\documentclass[%
  reprint,
  showpacs,preprintnumbers,
  amsmath,amssymb,
  aps,
  prb,
]{revtex4-2}

\usepackage{graphicx}[]
\usepackage{dcolumn}
\usepackage{bm}
\usepackage{multirow}
\usepackage{braket}
\usepackage{color}
\usepackage{ulem}

\usepackage{hyperref}
\usepackage{url}
\allowdisplaybreaks[4]

\begin{document}

\preprint{APS/123-QED}

\title{
Multiple-$q$ dipole-quadrupole instability in spin-1 triangular-lattice systems
}

\author{Satoru Hayami$^1$ and Kazumasa Hattori$^2$}
\affiliation{
$^1$Graduate School of Science, Hokkaido University, Sapporo 060-0810, Japan \\
$^2$Department of Physics, Tokyo Metropolitan University, Tokyo 192-0397, Japan
}
 
\begin{abstract}
We investigate a mechanism of multiple-$q$ states consisting of magnetic dipole and electric quadrupole degrees of freedom in spin-1 triangular-lattice systems. 
By systematically analyzing a minimal effective multipole model with bilinear multipole-multipole interactions in momentum space on a triangular lattice, we find minimum conditions of multipole interactions leading to the multiple-$q$ instability at zero temperature even without single-ion magnetic anisotropy and higher-order interactions. 
We present six types of triple-$q$ states with different magnetic dipole and electric quadrupole configurations depending on the model parameters. 
Our results indicate that the interplay between dipole and quadrupole degrees of freedom provides another source of inducing rich multiple-$q$ states. 
\end{abstract}
\maketitle

\section{Introduction}

Entanglement between orbital and spin degrees of freedom of electrons in solids has drawn considerable interest in condensed matter physics, since it becomes a fundamental origin of rich quantum states of matter. 
A concept of multipole has been introduced to describe atomic-scale entanglement in a unified way~\cite{Kusunose_JPSJ.77.064710, Santini_RevModPhys.81.807, kuramoto2009multipole}. 
So far, higher-rank multipole orderings have been identified mainly in $f$-electron systems, such as a rank-3 magnetic octupole in Ce$_{1-x}$La$_{x}$B$_6$~\cite{Mannix_PhysRevLett.95.117206} and rank-4 electric hexadecapole in PrRu$_4$P$_{12}$~\cite{lee2001structural,takimoto2006antiferro}. 
Among them, rank-2 electric quadrupole ordering is the most familiar multipole ordering. 
Indeed, such orderings have been observed not only in $f$-electron materials, such as CeB$_6$~\cite{Takigawa_doi:10.1143/JPSJ.52.728,nakao2001antiferro,tanaka2004direct,Portnichenko_PhysRevX.10.021010}, PrPb$_3$~\cite{morin1982magnetic, onimaru2004angle, Onimaru_PhysRevLett.94.197201}, and Pr$T_2$$X_{20}$ ($T=$ Ir, Rh, $X=$ Zn; $T=$ V, $X=$ Al)~\cite{Onimaru_PhysRevLett.106.177001, Ishii_doi:10.1143/JPSJ.80.093601, sakai2011kondo, onimaru2016exotic, Ishitobi_PhysRevB.104.L241110}, but also in $d$-electron materials as recently reported in Ba$_2$MgReO$_6$~\cite{Hirai_PhysRevResearch.2.022063, Mansouri_PhysRevMaterials.5.104410, Lovesey_PhysRevB.103.235160}. 

Recently, more exotic electronic orderings including the quadrupole degrees of freedom have been extensively studied. 
One of the examples is a CP$^2$ skyrmion consisting of a multiple-$q$ superposition of quadrupole density waves in a spin-1 model~\cite{Garaud_PhysRevB.87.014507, Akagi_PhysRevD.103.065008, Amari_PhysRevB.106.L100406, zhang2022cp}, which is different from a conventional magnetic skyrmion characterized by a CP$^1$ topological invariant without the quadrupole degree of freedom. 
Another example is different types of quadrupole orderings with different time-reversal and spatial inversion parities, such as magnetic quadrupole~\cite{Watanabe_PhysRevB.96.064432, thole2018magnetoelectric, Yanagi_PhysRevB.97.020404, Hayami_PhysRevB.104.045117}, electric toroidal quadrupole~\cite{Matteo_PhysRevB.96.115156, Hayami_PhysRevLett.122.147602, Ishitobi_doi:10.7566/JPSJ.88.063708}, and magnetic toroidal quadrupole~\cite{hayami2022spinconductivity}, as discussed in augmented multipole description~\cite{hayami2018microscopic, Hayami_PhysRevB.98.165110, Yatsushiro_PhysRevB.104.054412}. 
These unconventional quadrupole orderings are expected to exhibit qualitatively different physical phenomena from those found in conventional magnetic dipole (spin) orderings~\cite{kusunose2022generalization}. 
Thus, quadrupole systems provide a promising playground showing more intriguing electronic orderings and physical phenomena. 

In this study, we focus on a spin-1 system including magnetic dipole and electric quadrupole degrees of freedom to explore the possibility of multiple-$q$ multipole orderings~\cite{tsunetsugu2021quadrupole, hattori2022quadrupole, ishitobi2022triple}. 
We aim at elucidating unknown multiple-$q$ orderings as a consequence of the competition between the magnetic dipole and electric quadrupole moments.  
In order to clarify the relevant multipole interactions for the realization of such multiple-$q$ states, we analyze a set of minimal spin-1 models on a two-dimensional triangular lattice.
We systematically investigate the multiple-$q$ instability by analyzing fifteen variants of the spin-1 models with different multipole-multipole interactions. 
By performing the simulated annealing for the models, we find that four combinations of multipole interactions can induce a variety of triple-$q$ states with different magnetic dipole and electric quadrupole configurations. 
We show that the quadrupole degree of freedom becomes a source of rich multiple-$q$ states that have never been obtained in dipole-only systems.

The rest of this paper is organized as follows. 
In Sec.~\ref{sec: Setup}, we introduce the multipole moments in the spin-1 system and present an effective model with the bilinear multipole-multipole interactions in the momentum space on the triangular lattice. 
We also outline a numerical method based on the simulated annealing. 
In Sec.~\ref{sec: Results}, we show the phase diagrams under different multipole interactions, where we find that four combinations of multipole interactions exhibit different triple-$q$ states.  
We discuss the qualitative understanding of multiple-$q$ multipole orders, the effect of a hexagonal crystalline electric field, and a relationship with a charge disproportionation in Sec.~\ref{sec: Discussion}. 
Section~\ref{sec: Summary} is devoted to the summary of this paper.

\section{Setup}
\label{sec: Setup}

We focus on the multiple-$q$ instability in terms of the rank-1 magnetic dipole with three components $(M_x, M_y, M_z)$ and rank-2 electric quadrupole with five components $(Q_u, Q_v, Q_{yz}, Q_{zx}, Q_{xy})$ ($u= 3z^2-r^2$ and $v=x^2-y^2$) by considering an effective spin-1 ($S=1$) system. 
For an atomic basis set $\{\ket{\psi_x}, \ket{\psi_y}, \ket{\psi_z}\}$ with $\ket{\psi_x}=(\ket{S_z=1}+\ket{S_z=-1})/\sqrt{2}$, $\ket{\psi_y}=-i(\ket{S_z=1}-\ket{S_z=-1})/\sqrt{2}$, and $\ket{\psi_z}=\ket{S_z=0}$, where $S_z$ is the $z$ component of the effective $S=1$ state, the matrices for the three magnetic dipoles at the site $\bm{r}$ are given by
\begin{align}
\label{eq: Mxmat}
&M^{\bm{r}}_x=\left(
\begin{array}{ccc}
0 & 0 & 0 \\
0 & 0 & -i \\
0 & i & 0 
\end{array}
\right), \\
&M^{\bm{r}}_y=\left(
\begin{array}{ccc}
0 & 0 & i \\
0 & 0 & 0 \\
-i & 0 & 0 
\end{array}
\right), \\
&M^{\bm{r}}_z=\left(
\begin{array}{ccc}
0 & -i & 0 \\
i & 0 & 0 \\
0 & 0 & 0 
\end{array}
\right), 
\end{align}
and those for the five electric quadrupoles are given by 
\begin{align}
&Q^{\bm{r}}_u=\frac{1}{\sqrt{3}}\left(
\begin{array}{ccc}
-1 & 0 & 0 \\
0 & -1 & 0 \\
0 & 0 & 2 
\end{array}
\right), \label{eq: QuMat}\\
&Q^{\bm{r}}_v=\left(
\begin{array}{ccc}
1 & 0 & 0 \\
0& -1 & 0 \\
0 & 0 & 0 
\end{array}
\right), \\
&Q^{\bm{r}}_{yz}=\left(
\begin{array}{ccc}
0 & 0 & 0 \\
0& 0 & 1 \\
0 & 1 & 0 
\end{array}
\right),\\
&Q^{\bm{r}}_{zx}=\left(
\begin{array}{ccc}
0 & 0 & 1 \\
0& 0 & 0 \\
1 & 0 & 0 
\end{array}
\right), \\
\label{eq: Qxymat}
&Q^{\bm{r}}_{xy}=\left(
\begin{array}{ccc}
0 & 1 & 0 \\
1 & 0 & 0 \\
0 & 0 & 0 
\end{array}
\right). 
\end{align}
As we have adopted the real-number representation of the basis wave function, the magnetic dipoles (electric quadrupoles) are expressed as the imaginary (real) matrix owing to the time-reversal parity. 
All the multipoles are invariant under the spatial inversion symmetry.

\begin{figure}[t!]
\begin{center}
\includegraphics[width=0.6 \hsize ]{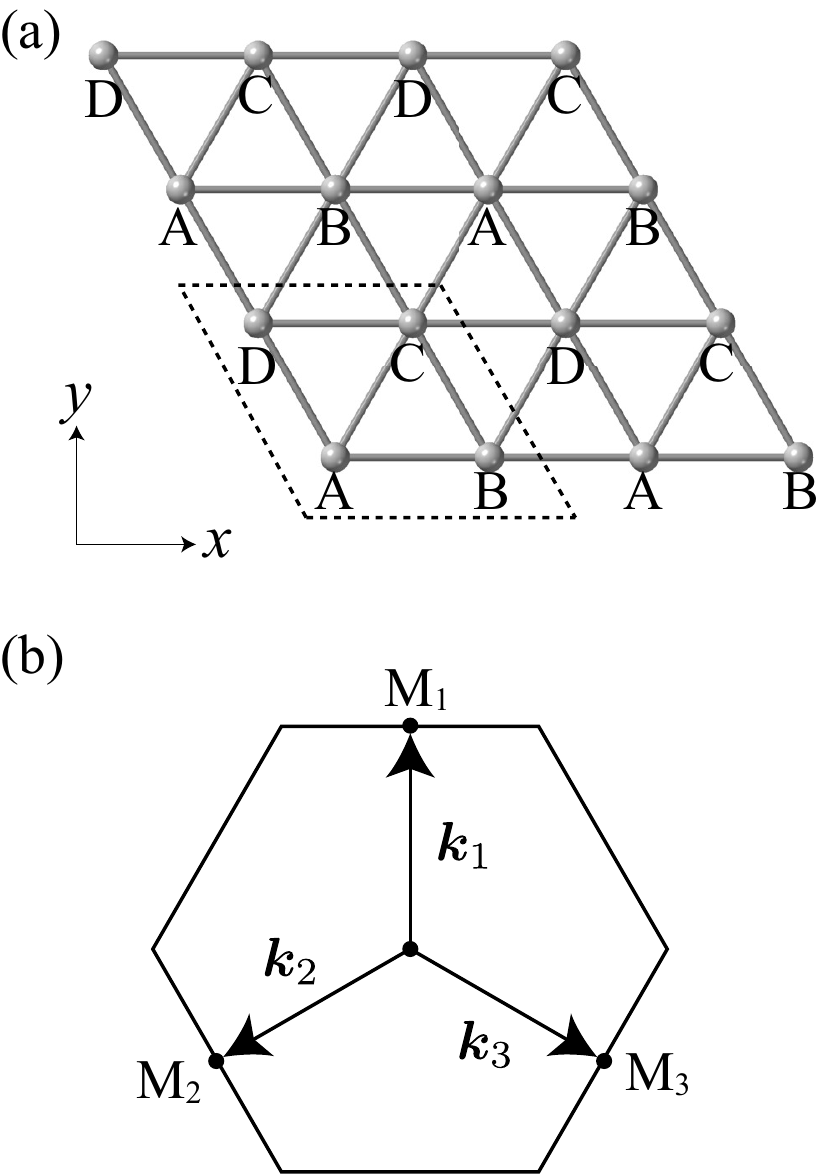} 
\caption{
\label{fig: lattice}
(a) Triangular lattice with four sublattices A--D. 
The dashed parallelogram stands for the magnetic unit cell. 
(b) The first Brillouin zone with the triple-$q$ ordering vectors $\bm{k}_1$, $\bm{k}_2$, $\bm{k}_3$ at the M$_1$, M$_2$, M$_3$ points, respectively. 
}
\end{center}
\end{figure}

To discuss the multiple-$q$ instability, we analyze a localized model including these multipole degrees of freedom on a two-dimensional triangular lattice under the point group $D_{\rm 6h}$ [Fig.~\ref{fig: lattice}(a)], where $(M_x^{\bm{r}}, M_y^{\bm{r}})$, $M_z^{\bm{r}}$, $Q_u^{\bm{r}}$, $(Q^{\bm{r}}_{yz}, Q^{\bm{r}}_{zx})$, and $(Q^{\bm{r}}_v, Q^{\bm{r}}_{xy})$ belong to the $\Gamma^-_5$, $\Gamma^-_2$, $\Gamma^+_1$, $\Gamma^+_5$, and $\Gamma^+_6$ irreducible representations, respectively; the superscript $+ (-)$ represents the time-reversal parity $+1$ $(-1)$. 
It is noted that $Q_u^{\bm{r}}$ belongs to the totally symmetric irreducible representation. 
We set the lattice constant to unity. 

Then, the model Hamiltonian including minimal multipole-multipole interactions is generally given by
\begin{align}
\label{eq: Ham_real}
\mathcal{H}= -&\sum_{\bm{r}\bm{r}'} \{
J^{\Gamma_5(\bm{r}\bm{r}')}_{\rm M} \left(M^{\bm{r}}_xM^{\bm{r}'}_x+M^{\bm{r}}_y M^{\bm{r}'}_y\right) \nonumber \\ 
&+ J^{\Gamma_2 (\bm{r}\bm{r}')}_{\rm M}M^{\bm{r}}_z M^{\bm{r}'}_z +J^{\Gamma_1 (\bm{r}\bm{r}')}_{\rm Q} Q^{\bm{r}}_u Q^{\bm{r}'}_u \nonumber \\
& +J^{\Gamma_5 (\bm{r}\bm{r}')}_{\rm Q} \left(Q^{\bm{r}}_{yz} Q^{\bm{r}'}_{yz} +Q^{\bm{r}}_{zx} Q^{\bm{r}'}_{zx} \right) \nonumber \\ 
&+J^{\Gamma_6 (\bm{r}\bm{r}')}_{\rm Q} \left(Q^{\bm{r}}_{v} Q^{\bm{r}'}_{v} +Q^{\bm{r}}_{xy} Q^{\bm{r}'}_{xy} \right)\}
+ \Delta \sum_{\bm{r}} Q^{\bm{r}}_u,
\end{align}
where we fix the moment length at each lattice site as 
$\sum_X (X^{\bm{r}})^2 = 4/3$ for $X=M_x, M_y, M_z, Q_u, Q_v, Q_{yz}, Q_{zx}, Q_{xy}$. 
The first term represents multipole-multipole interactions. 
We set five exchange interaction constants $(J^{\Gamma_2 (\bm{r}\bm{r}')}_{\rm M}, J^{\Gamma_5 (\bm{r}\bm{r}')}_{\rm M}, J^{\Gamma_1 (\bm{r}\bm{r}')}_{\rm Q}, J^{\Gamma_5 (\bm{r}\bm{r}')}_{\rm Q}, J^{\Gamma_6 (\bm{r}\bm{r}')}_{\rm Q})$ between the sites at $\bm{r}$ and $\bm{r}'$ according to the irreducible representation under the point group $D_{\rm 6h}$. 
The superscript in the interaction constants stands for the irreducible representation; we take the diagonal-type interaction between the same multipole. 
The last term represents the onsite potential that arises from the hexagonal crystalline electric field. 
It is noted that the model (\ref{eq: Ham_real}) is an extension of the spin-1 bilinear-biquadratic model~\cite{Fath_PhysRevB.51.3620, Schollwock_PhysRevB.53.3304, Harada_PhysRevB.65.052403, Lauchli_PhysRevLett.97.087205, tsunetsugu2006spin, tsunetsugu2007spin, Smerald_PhysRevB.88.184430}; we set the different interaction constants for different multipoles from the symmetry viewpoint.

For the model (\ref{eq: Ham_real}), we focus on the multiple-$q$ instability at the M$_1$--M$_3$ points, which are located at the boundary of the Brillouin zone, i.e., $\bm{k}_1=(0, K_{\rm M})$, $\bm{k}_2=(-K_{\rm M}/2,-\sqrt{3}K_{\rm M}/2)$, and $\bm{k}_3=(-K_{\rm M}/2,\sqrt{3}K_{\rm M}/2)$ with $K_{\rm M}=2\pi/\sqrt{3}$, as shown in Fig.~\ref{fig: lattice}(b). 
In other words, we consider the four-sublattice structure with the sublattices A--D, as shown in Fig.~\ref{fig: lattice}(a). 
The instability at the M points can be achieved by considering the competition between the ferroic nearest-neighbor interaction and the antiferroic second-nearest-neighbor interaction in the real space~\cite{Momoi_PhysRevLett.79.2081} or the Ruderman-Kittel-Kasuya-Yosida interaction~\cite{Ruderman, Kasuya, Yosida1957} by the Fermi surface nesting at particular electron fillings in the itinerant electron model~\cite{Martin_PhysRevLett.101.156402, Akagi_JPSJ.79.083711, Akagi_PhysRevLett.108.096401, Hayami_PhysRevB.90.060402, hayami_PhysRevB.91.075104, hayami2021topological}.
Especially for the latter case, the coupling constants in different multipole channels can be comparable with each other~\cite{ohkawa1983ordered, shiina1997magnetic, yamada2019derivation}. 
Assuming that such interactions are dominantly important in the present system, one can consider an effective spin model with the momentum-resolved interaction as follows: 
\begin{align}
\label{eq: Ham}
\mathcal{H}= -&\sum_{\bm{q}=\bm{k}_1, \bm{k}_2, \bm{k}_3} \{
J^{\Gamma_5}_{\rm M} \left[M_x(\bm{q})M_x(-\bm{q})+M_y(\bm{q})M_y(-\bm{q})\right]\nonumber \\
&+ J^{\Gamma_2}_{\rm M}M_z(\bm{q})M_z(-\bm{q})+J^{\Gamma_1}_{\rm Q} Q_u(\bm{q})Q_u(-\bm{q}) \nonumber \\
&+J^{\Gamma_5}_{\rm Q} \left[Q_{yz}(\bm{q})Q_{yz}(-\bm{q}) +Q_{zx}(\bm{q})Q_{zx}(-\bm{q}) \right] \nonumber \\ 
&+J^{\Gamma_6}_{\rm Q} \left[Q_{v}(\bm{q})Q_{v}(-\bm{q}) +Q_{xy}(\bm{q})Q_{xy}(-\bm{q}) \right]\} \nonumber \\
&+ \Delta \sum_{\bm{r}} Q^{\bm{r}}_u,
\end{align}
where $X(\bm{q})$ is the Fourier transform of $X^{\bm{r}}$; $(J^{\Gamma_2}_{\rm M}, J^{\Gamma_5}_{\rm M}, J^{\Gamma_1}_{\rm Q}, J^{\Gamma_5}_{\rm Q}, J^{\Gamma_6}_{\rm Q})$ represent the interaction constants at the M points; $X(\bm{q})$ at $\bm{q}=\bm{k}_1$, $\bm{k}_2$, and $\bm{k}_3$ is real number as their momenta are time-reversal invariant one. 
Although the magnitudes of $(J^{\Gamma_2}_{\rm M}, J^{\Gamma_5}_{\rm M}, J^{\Gamma_1}_{\rm Q}, J^{\Gamma_5}_{\rm Q}, J^{\Gamma_6}_{\rm Q})$ are related to $(J^{\Gamma_2 (\bm{r}\bm{r}')}_{\rm M}, J^{\Gamma_5 (\bm{r}\bm{r}')}_{\rm M}, J^{\Gamma_1 (\bm{r}\bm{r}')}_{\rm Q}, J^{\Gamma_5 (\bm{r}\bm{r}')}_{\rm Q}, J^{\Gamma_6 (\bm{r}\bm{r}')}_{\rm Q})$, we take them as phenomenological parameters for simplicity. 
In addition, we neglect the interactions at the other wave vectors, since they are irrelevant in determining the low-temperature phase diagram.
Hereafter, we use $\sideset{}{'}{\sum}_{\bm{q}}$ instead of $\sum_{\bm{q}=\bm{k}_1, \bm{k}_2, \bm{k}_3}$ for notational simplicity.

We calculate the lowest-energy four-sublattice multipole configurations of the model (\ref{eq: Ham}) by performing the simulated annealing for the four-sublattice structure ($\bm{r}$: A--D) under the periodic boundary condition. 
In the simulations, the local updates for localized multipoles are done, based on the Metropolis algorithm.
By using four variables $\theta$, $\psi$, $\alpha_1$, and $\alpha_2$ and a coherent state for SU(3) as $e^{i \alpha_1}\sin \theta \cos \phi \ket{\psi_x} + e^{i \alpha_2} \sin \theta \sin \phi \ket{\psi_y} + \cos \theta \ket{\psi_z}$~\cite{mathur2001coherent, Zhang_PhysRevB.104.104409}, the multipoles (\ref{eq: Mxmat})--(\ref{eq: Qxymat}) are evaluated as 
\begin{align}
\label{eq: Mx_coh}
M_x&= -\sin \alpha_2 \sin 2\theta \sin \phi,  \\
\label{eq: My_coh}
M_y&= \sin \alpha_1 \sin 2\theta \cos \phi,  \\
\label{eq: Mz_coh}
M_z&=-\sin(\alpha_1-\alpha_2)\sin^2 \theta \sin 2 \phi, \\
\label{eq: Qu_coh}
Q_u&=\frac{1}{2\sqrt{3}}(1+3\cos 2\theta), \\
\label{eq: Qv_coh}
Q_v&= \sin^2 \theta \cos 2\phi, \\
\label{eq: Qyz_coh}
Q_{yz}&= \cos \alpha_2 \sin 2\theta \sin \phi, \\
\label{eq: Qzx_coh}
Q_{zx}&= \cos \alpha_1 \cos \phi \sin 2\theta, \\
\label{eq: Qxy_coh}
Q_{xy}&= \cos (\alpha_1 - \alpha_2) \sin^2 \theta \sin 2 \phi, 
\end{align}
where $0 \leq \theta, \phi \leq \pi/2$ and $0 \leq \alpha_1, \alpha_2 \leq 2\pi$. 
The temperature is gradually reduced from the initial temperature $T_0$ to the final lowest temperature $T_f$ with the annealing rate $\alpha = 0.999999$ in each Monte Carlo step, where $T_0$ and $T_f$ are typically set as $T_0=1$ and $T_f=0.001$, respectively. 
When the temperature reaches $T_f$, we perform $10^5$--$10^6$ Monte Carlo sweeps for equilibration and measurements. 
The above process is independently done in each set of model parameters. 
Although we start the simulations by using a random multipole configuration, we also start the simulations from the multipole configurations obtained at low temperatures. 
When we obtain several solutions in the same model parameter, we adopt the lowest-energy state.

\section{Results}
\label{sec: Results}

We examine the multiple-$q$ dipole-quadrupole instability in the model (\ref{eq: Ham}). 
To systematically investigate what types of multipole interactions result in the multiple-$q$ states, we consider various situations by setting one or two out of the five interaction constants $(J^{\Gamma_2}_{\rm M}, J^{\Gamma_5}_{\rm M}, J^{\Gamma_1}_{\rm Q}, J^{\Gamma_5}_{\rm Q}, J^{\Gamma_6}_{\rm Q})$ to be nonzero. 
For systems with more than two multipole moments, the situation becomes more complex.
However, by focusing on the models with two multipole moments, we can clarify the inter-multipole correlations and their characteristics.
We take $\Delta=0$ in this section. We discuss the effect of $\Delta$ in Sec.~\ref{sec: Discussion}.

\begin{table}[tb!]
\caption{
Summary of the multiple-$q$ instabilities for the effective dipole-quadrupole model (\ref{eq: Ham}) when one or two out of five interaction constants are finite. The column and the raw represent the irreducible representation of these multipole moments in $D_{\rm 6h}$. The rightmost column represents the notation of multipoles belonging to each irreducible representation. The diagonal elements represent the ground state when only one interaction parameter is present between the corresponding multipole and they are single-$q$ state (1$q$) for all the cases. For off-diagonal ones, they correspond to the model with two interaction parameters in the two multipoles.
3$q$-I and 3$q$-II represent different types of the triple-$q$ states.  
}
\label{tab: multiQ}
\centering
\renewcommand{\arraystretch}{1.2}
 \begin{tabular}{lcccccc}
 \hline  \hline
& $\Gamma^+_{1}$ & $\Gamma^+_{5}$ & $\Gamma^+_{6}$ & $\Gamma^-_{2}$ & $\Gamma^-_{5}$ & multipole \\ \hline 
$\Gamma^+_{1} $  & 1$q$ & 1$q$ & 3$q$-I & $3q$-I & 1$q$ & $Q_u$\\ 
$\Gamma^+_{5}$ & & $1q$ & $3q$-II & $1q$ & $1q$ & $Q_{yz}, Q_{zx}$\\ 
$\Gamma^+_{6} $ & & & 1$q$& 1$q$ & 3$q$-II  & $Q_v, Q_{xy}$ \\ 
$\Gamma^-_{2}$ & & & & 1$q$ & 1$q$ & $M_z$\\ 
$\Gamma^-_{5}$ &  & &  & & 1$q$ & $M_x, M_y$ \\ 
 \hline
\hline 
\end{tabular}
\end{table}

First, we suppose five situations where only one interaction constant is nonzero and $\Delta=0$. 
In this case, the single-$q$ state becomes the ground state irrespective of the interaction types. 
It is noted that the double-$q$ state has the same energy as the single-$q$ state does in the models with the two-dimensional irreducible representations, $(M_x, M_y)$, $(Q_{yz}, Q_{zx})$, and $(Q_{v}, Q_{xy})$. 
Next, we set two out of the five interactions to nonzero while keeping $\Delta=0$; there are ten such combinations. 
We find that the triple-$q$ instabilities occur for the four cases: 
$J^{\Gamma_1}_{\rm Q}$--$J^{\Gamma_6}_{\rm Q}$, $J^{\Gamma_1}_{\rm Q}$--$J^{\Gamma_2}_{\rm M}$, $J^{\Gamma_5}_{\rm Q}$--$J^{\Gamma_6}_{\rm Q}$, and $J^{\Gamma_6}_{\rm Q}$--$J^{\Gamma_5}_{\rm M}$. 
We can classify the obtained triple-$q$ states into two types according to the spatial distribution of quadrupole moments: 3$q$-I and 3$q$-II, as summarized in Table~\ref{tab: multiQ}, where 3$q$-I exhibits different amplitudes of quadrupole moments in each sublattice, while 3$q$-II shows the same amplitude of quadrupole moments in all the sublattices.
Reflecting its difference, the former accompanies the charge disproportionation in the presence of itinerant electrons, while the latter does not; see Sec.~\ref{sec: Charge disproportionation} for details.
In the following, we discuss the ground-state phase diagram for the four models: 
the $J^{\Gamma_1}_{\rm Q}$--$J^{\Gamma_2}_{\rm M}$ model in Sec.~\ref{sec: model1} and the $J^{\Gamma_1}_{\rm Q}$--$J^{\Gamma_6}_{\rm Q}$ model in Sec.~\ref{sec: model2} categorized into the 3$q$-I, and the $J^{\Gamma_5}_{\rm Q}$--$J^{\Gamma_6}_{\rm Q}$ model in Sec.~\ref{sec: model3} and the $J^{\Gamma_6}_{\rm Q}$--$J^{\Gamma_5}_{\rm M}$ model in Sec.~\ref{sec: model4} categorized into the 3$q$-II.

\subsection{$J^{\Gamma_1}_{\rm Q}$--$J^{\Gamma_2}_{\rm M}$ model}
\label{sec: model1}

The $J^{\Gamma_1}_{\rm Q}$--$J^{\Gamma_2}_{\rm M}$ model is given by
\begin{align}
\label{eq: Ham1}
\mathcal{H}= &-\sideset{}{'}{\sum}_{\bm{q}} \{
 J^{\Gamma_2}_{\rm M}M_z(\bm{q})M_z(-\bm{q})+J^{\Gamma_1}_{\rm Q} Q_u(\bm{q})Q_u(-\bm{q})\},
\end{align}
where $J^{\Gamma_5}_{\rm M}= J^{\Gamma_5}_{\rm Q}=J^{\Gamma_6}_{\rm Q}=0$. 
To investigate the phase diagram of the model with two independent model parameters, we set $ J^{\Gamma_2}_{\rm M}=J \cos \theta_{\rm I}$ and $J^{\Gamma_1}_{\rm Q}=J \sin \theta_{\rm I}$ with $J=1$ and $0 \leq \theta_{\rm I} \leq \pi/2$; $J$ is the energy unit of the model. 

\begin{figure}[t!]
\begin{center}
\includegraphics[width=1.0 \hsize ]{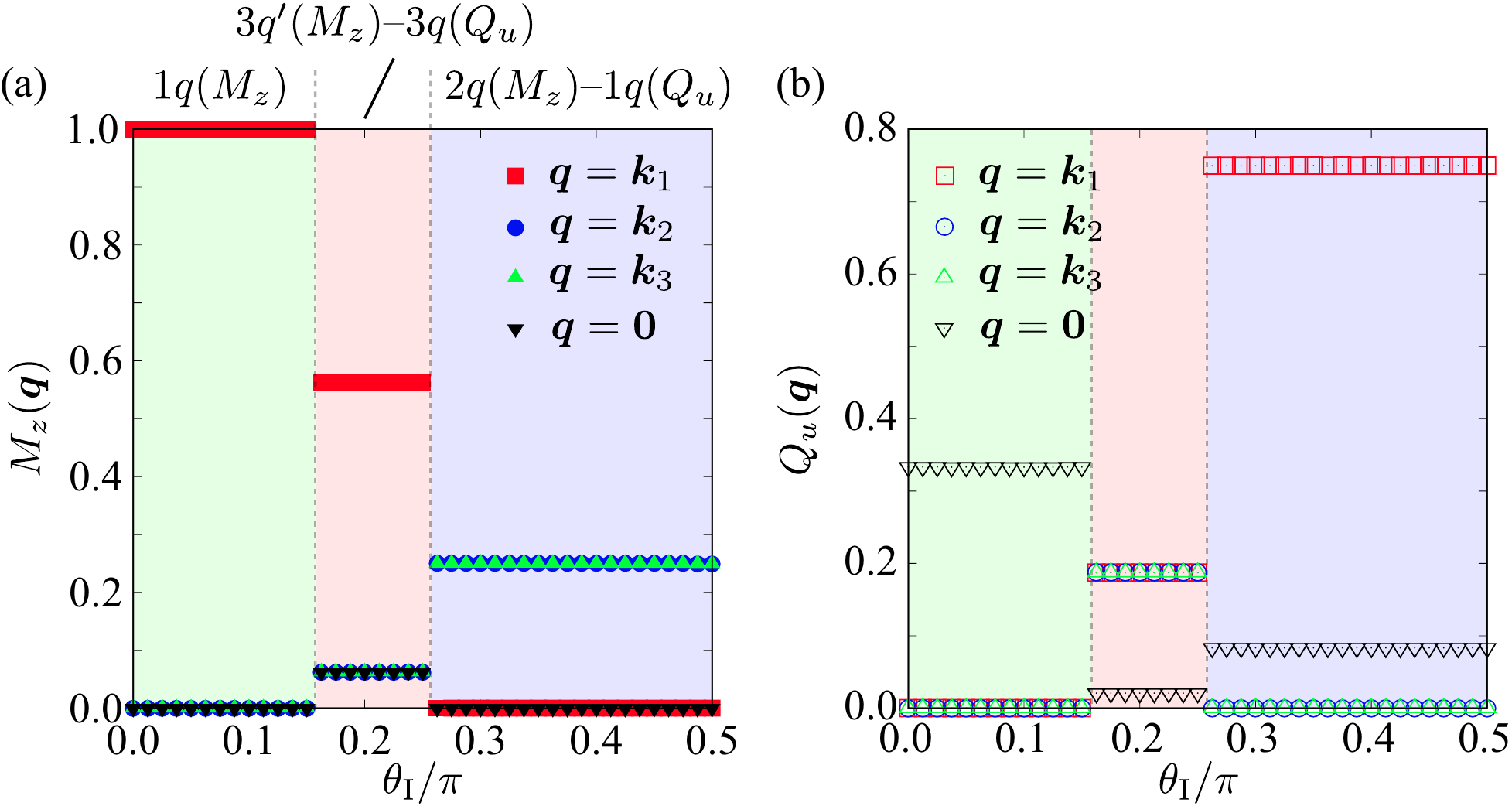} 
\caption{
\label{fig: QuMz}
$\theta_{\rm I}$ dependence of (a) $[M_z(\bm{q})]^2$ and (b) $[Q_u(\bm{q})]^2$ at $\bm{q}=\bm{k}_1$, $\bm{k}_2$, $\bm{k}_3$, and $\bm{0}$ for the model (\ref{eq: Ham1}). 
The vertical dashed lines represent the phase boundaries. 
}
\end{center}
\end{figure}

\begin{figure}[t!]
\begin{center}
\includegraphics[width=1.0 \hsize ]{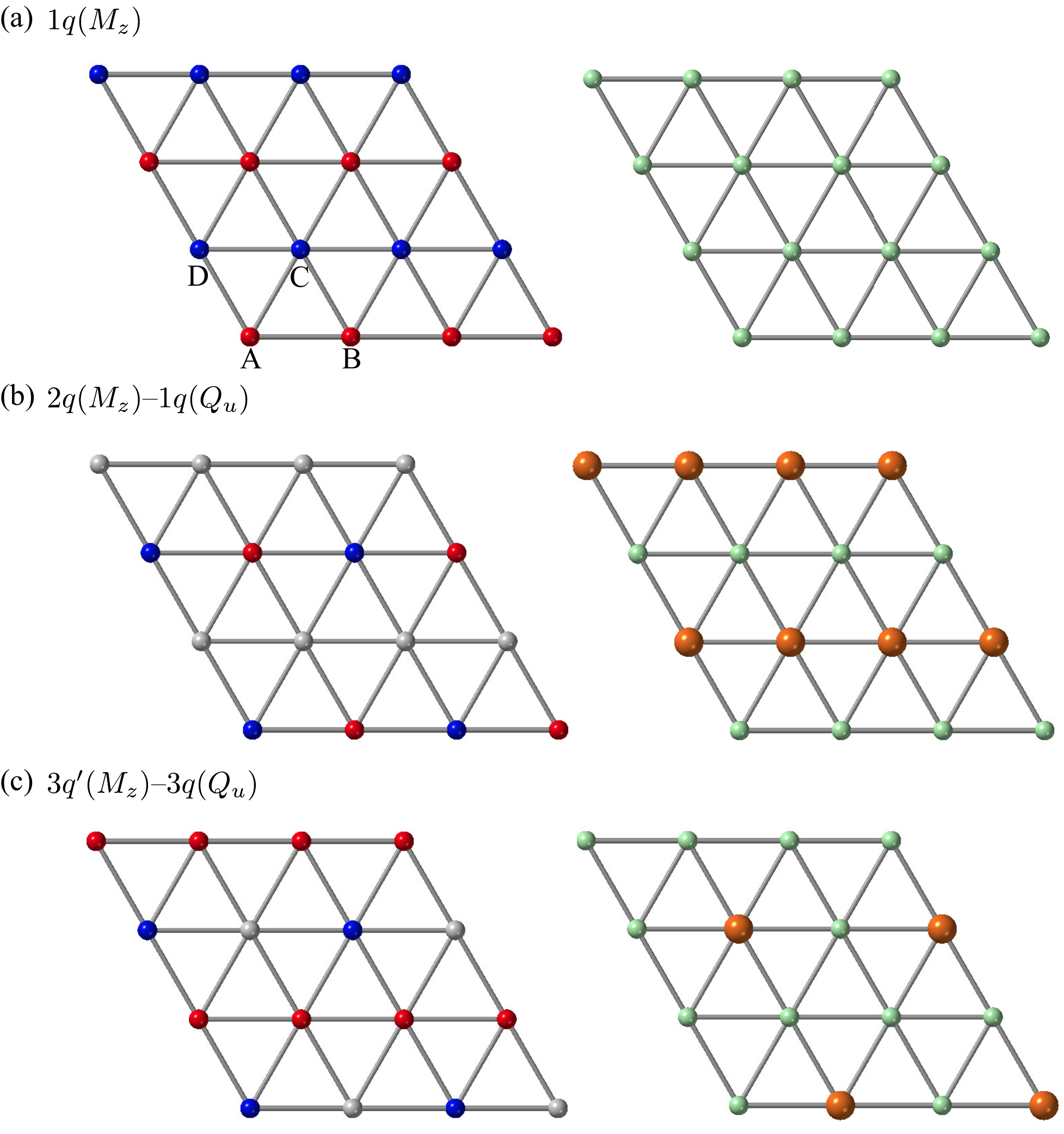} 
\caption{
\label{fig: QuMz_RS}
Schematic real-space configurations of (left panel) $M^{\bm{r}}_z$ and (right panel) $Q^{\bm{r}}_u$ for (a) the 1$q(M_z)$, (b) $2q(M_z)$--$1q(Q_u)$, and (c) $3q'(M_z)$--$3q(Q_u)$ phases. 
In the left panel, the red (blue) spheres represent $M^{\bm{r}}_z=1$ ($M^{\bm{r}}_z=-1$). 
In the right panel, the orange (green) spheres represent $Q^{\bm{r}}_u=2/\sqrt{3}$ ($Q^{\bm{r}}_u=-1/\sqrt{3}$). 
}
\end{center}
\end{figure}

Figures~\ref{fig: QuMz}(a) and \ref{fig: QuMz}(b) show the $\theta_{\rm I}$ dependence of $[M_z(\bm{q})]^2$ and $[Q_u(\bm{q})]^2$ for $\bm{q}=\bm{k}_1$, $\bm{k}_2$, $\bm{k}_3$, and $\bm{0}$, which are obtained by the simulated annealing.  
In the simulations, the symmetry-equivalent states were obtained with equal probability through several runs with different random initial configurations. 
Thus, we show the data in each ordered state by appropriately sorting $[M_z(\bm{q})]^2$ and $[Q_u(\bm{q})]^2$ for better readability. 
There are three phases while $\theta_{\rm I}$ is varied, which are denoted as the 1$q (M_z)$, 3$q'(M_z)$--3$q(Q_u)$, and 2$q(M_z)$--1$q(Q_u)$ states from small to large $\theta_{\rm I}$; $3q(X)$ means the isotropic triple-$q$ state with $X(\bm{k}_1)=X(\bm{k}_2)=X(\bm{k}_3)$, while 3$q'(X)$ means the anisotropic triple-$q$ state with the different amplitudes of $X(\bm{q})$ at $\bm{q}=\bm{k}_1$, $\bm{k}_2$, and $\bm{k}_3$. 
The phase transitions among these phases are characterized by the first-order transition with jumps in $M_z(\bm{q})$ and $Q_u(\bm{q})$.

For small $\theta_{\rm I}$, i.e., $J^{\Gamma_2}_{\rm M} > J^{\Gamma_1}_{\rm Q}$, the 1$q(M_z)$ state with $M_z(\bm{k}_1)=1$ is stabilized so as to gain the energy by $J^{\Gamma_2}_{\rm M}$. 
In the real space, $(M^{\rm A}_z, M^{\rm B}_z, M^{\rm C}_z, M^{\rm D}_z)=(+1, +1, -1, -1)$, as shown in the left panel of Fig.~\ref{fig: QuMz_RS}(a). 
In other words, $\theta=\pi/2$ and $\phi=\pi/4$ for all the sublattices and $\alpha_1=\alpha_2-\pi/2$ ($\alpha_1=\alpha_2+\pi/2$) for the sublattices A and B (C and D). See Eq.~(\ref{eq: Mz_coh}). 
One also finds that $Q_u^{\bm{r}} = -1/\sqrt{3}$ at all the sublattices, and the other multipoles in Eqs.~(\ref{eq: Mx_coh})--(\ref{eq: Qxy_coh}) are zero. 
As shown in the right panel of Fig.~\ref{fig: QuMz_RS}(a), the uniform component of $Q^{\bm{r}}_u$ remains as $[Q_u(\bm{0})]^2=1/3$.

For large $\theta_{\rm I}$, i.e., $J^{\Gamma_2}_{\rm M} < J^{\Gamma_1}_{\rm Q}$, the 2$q(M_z)$--1$q(Q_u)$ state with $[Q_u(\bm{k}_1)]^2=3/4$, $[M_z(\bm{k}_2)]^2=[M_z(\bm{k}_3)]^2=1/4$, and $[Q_u(\bm{0})]^2=1/12$ is realized, as shown in Figs.~\ref{fig: QuMz}(a) and \ref{fig: QuMz}(b). 
In contrast to the 1$q(M_z)$ state, this state is characterized by not only the single-$q$ component of $Q_u(\bm{k}_\eta)$ ($\eta=1$--3) but also the double-$q$ component of $M_z(\bm{k}_{\eta'})$ with $\eta' \neq \eta$. 
Thus, this is a mixed triple-$q$ state with two different multipole moments. 
This result means that the additional double-$q$ modulation with $M_{z}(\bm{k}_2)=M_{z}(\bm{k}_3)$ to the single-$q$ state with $Q_{u}(\bm{k}_1)$ leads to the energy gain by introducing infinitesimally small $J^{\Gamma_2}_{\rm M}$. 

The double-$q$ feature regarding $M_z$ is understood from the local constraint of the multipole length as $\sum_X (X^{\bm{r}})^2 = 4/3$ in the real-space picture. 
As $J^{\Gamma_2}_{\rm M} < J^{\Gamma_1}_{\rm Q}$, the single-$q$ state with $Q_u (\bm{k}_1)$ is realized to gain the energy by $J^{\Gamma_1}_{\rm Q}$ as much as possible, which leads to $(Q^{\rm A}_u, Q^{\rm B}_u, Q^{\rm C}_u, Q^{\rm D}_u)=(-1/\sqrt{3}, -1/\sqrt{3}, 2/\sqrt{3}, 2/\sqrt{3})$, as shown in the right panel of Fig.~\ref{fig: QuMz_RS}(b). 
Thus, $\theta=0$ for the sublattices C and D, which means that the other multipoles vanish.
Meanwhile, $\theta=\pi/2$ for the sublattices A and B; 
there are internal degrees of freedom in terms of $\phi$, $\alpha_1$, and $\alpha_2$, which are determined to gain the energy by $J^{\Gamma_2}_{\rm M}$; $M^{\bm{r}}_z$ becomes maximum by setting $\phi=\pi/4$ and $\alpha_1=\alpha_2 \mp \pi/2$ corresponding to $M^{\bm{r}}_z=\pm 1$. 
Since the alignment of $(M^{\rm A}_z, M^{\rm B}_z)=(-1, +1)$ with $[M_{z}(\bm{k}_2)]^2=[M_{z}(\bm{k}_3)]^2=1/4$ gives the lower energy than that of $(M_z^{\rm A}, M_z^{\rm B})=(+1, +1)$ with $[M_{z}(\bm{k}_1)]^2=[M_{z}(\bm{0})]^2=1/4$, the double-$q$ structure of $M_z(\bm{q})$ is obtained. 
The real-space alignment of $M^{\bm{r}}_z$ is shown in the left panel of Fig.~\ref{fig: QuMz_RS}(b). 
The coexistence of the single-$q$ modulation in terms of $Q_u(\bm{q})$ and double-$q$ modulation in terms of $M_z(\bm{q})$ is also reasonable from the viewpoint of symmetry; an effective coupling in the form of $Q_u(\bm{k}_1)M_z(\bm{k}_2)M_z(\bm{k}_3)$ belongs to the totally symmetric irreducible representation. 
As for $Q_u (\bm{0})$, the relevant coupling is given as $Q_u (\bm{0})Q_u(\bm{k}_1)Q_u(-\bm{k}_1)$, $Q_u (\bm{0})M_z(\bm{k}_2)M_z(-\bm{k}_2)$, and $Q_u (\bm{0})M_z(\bm{k}_3)M_z(-\bm{k}_3)$. 
Since such couplings usually appear in the free energy, it is expected that the obtained state can remain stable against thermal fluctuations~\cite{tsunetsugu2021quadrupole, hattori2022quadrupole, ishitobi2022triple}.

From the above argument, one notices that a similar triple-$q$ structure is possible for the $J^{\Gamma_1}_{\rm Q}$--$J^{\Gamma_6}_{\rm Q}$ model, as discussed in Sec.~\ref{sec: model2}, but impossible for the $J^{\Gamma_1}_{\rm Q}$--$J^{\Gamma_5}_{\rm Q}$ and $J^{\Gamma_1}_{\rm Q}$--$J^{\Gamma_5}_{\rm M}$ models. 
This is because the 1$q$ modulation of $Q_u(\bm{q})$ is expressed as $\theta=0$ or $\theta=\pi/2$, i.e., $M^{\bm{r}}_x=M^{\bm{r}}_y=Q^{\bm{r}}_{yz}=Q^{\bm{r}}_{zx}=0$ in Eqs.~(\ref{eq: Mx_coh}), (\ref{eq: My_coh}), (\ref{eq: Qyz_coh}), and (\ref{eq: Qzx_coh}). 

In the intermediate $\theta_{\rm I}$, another triple-$q$ state denoted as the 3$q'(M_z)$--3$q(Q_u)$ state appears. 
This state exhibits the anisotropic triple-$q$ structure of $M_z(\bm{q})$ in Fig.~\ref{fig: QuMz}(a) and the isotropic one of $Q_u(\bm{q})$ in Fig.~\ref{fig: QuMz}(b), which are qualitatively different from those in the 1$q(M_z)$ and 2$q(M_z)$--1$q(Q_u)$ states. 
The latter isotropic triple-$q$ structure of $Q_u(\bm{q})$ leads to the disproportionation of $Q^{
{\bm{r}}}_u$ in the real space with the ratio 3:1, as shown in the right panel of Fig.~\ref{fig: QuMz_RS}(c); $Q_u^{\bm{r}}$ at the sublattice B takes $Q_u^{\rm B}=2/\sqrt{3}$, while those at the other three sublattices take $Q_u^{\bm{r}}=-1/\sqrt{3}$. 
This alignment of $Q_u^{\bm{r}}$ is regarded as a spontaneous kagome network consisting of the sublattices A, C, and D. 
Then, $M^{\bm{r}}_z$ is aligned on these sublattices so as to gain the energy by $J^{\Gamma_2}_{\rm M}$, which results in the up-up-down structure on the kagome network, as shown in the left panel of Fig.~\ref{fig: QuMz_RS}(c). 
This consists of $M_z(\bm{k}_1)$ with imposing $M^{\bm{r}}_z=0$ at the center of the kagome hexagons (sublattice B), where $Q^{\rm B}_u=2/\sqrt{3}$.
Thus, the amplitude of $M_z(\bm{k}_1)$ is larger than those of $M_z(\bm{k}_2)=M_z(\bm{k}_3)$. 
Similarly to the 2$q(M_z)$--1$q(Q_u)$ state, the coexistence of the isotropic triple-$q$ modulation in terms of $Q_u(\bm{q})$ and anisotropic triple-$q$ modulation in terms of $M_z(\bm{q})$ in the 3$q'(M_z)$--3$q(Q_u)$ state is also reasonable from the symmetry; the effective coupling $c [Q_u(\bm{k}_1)M_z(\bm{k}_2)M_z(\bm{k}_3)+Q_u(\bm{k}_2)M_z(\bm{k}_3)M_z(\bm{k}_1)+Q_u(\bm{k}_3)M_z(\bm{k}_1)M_z(\bm{k}_2)]$ belongs to the totally symmetric irreducible representation, which contributes to the free energy. 
When the coefficient $c$ is positive and $Q_u(\bm{k}_1)=Q_u(\bm{k}_2)=Q_u(\bm{k}_3)>0$, this effective coupling favors the anisotropic triple-$q$ modulation of $M_z(\bm{q})$ with $2M_z(\bm{k}_1)=-M_z(\bm{k}_2)=-M_z(\bm{k}_3)$. 

As shown in Figs.~\ref{fig: QuMz}(a) and \ref{fig: QuMz}(b), the moments of $M_z(\bm{q})$ and $Q_{u}(\bm{q})$ are independent of $\theta_{\rm I}$ inside each phase. 
Thus, one can analytically determine the phase boundaries by evaluating the internal energy in each phase, which are given by
\begin{align}
E^{1q(M_z)} &= -J^{\Gamma_2}_{\rm M}, \\
E^{3q'(M_z)-3q(Q_u)} &=-\frac{11}{16}J^{\Gamma_2}_{\rm M}-\frac{9}{16}J^{\Gamma_1}_{\rm Q}, \label{eq:E3qMz-3qQu} \\
E^{2q(M_z)-1q(Q_u)} &= -\frac{1}{2}J^{\Gamma_2}_{\rm M} - \frac{3}{4}J^{\Gamma_1}_{\rm Q}, 
\end{align}
where $E^{1q(M_z)}$, $E^{3q'(M_z)-3q(Q_u)}$, and $E^{2q(M_z)-1q(Q_u)}$ represent the energy for the 1$q (M_z)$, 3$q'(M_z)$--3$q(Q_u)$, and 2$q(M_z)$--1$q(Q_u)$ states, respectively. 
By comparing the energies, the phase boundary between the 1$q (M_z)$ and 3$q'(M_z)$--3$q(Q_u)$ states are given by $\arctan(5/9)\sim 0.161 \pi$ and that between the 3$q'(M_z)$--3$q(Q_u)$ and 2$q(M_z)$--1$q(Q_u)$ states is given by $\theta=0.25 \pi$, which are consistent with the numerical results in Figs.~\ref{fig: QuMz}(a) and \ref{fig: QuMz}(b).

\subsection{$J^{\Gamma_1}_{\rm Q}$--$J^{\Gamma_6}_{\rm Q}$ model}
\label{sec: model2}

Next, we consider the $J^{\Gamma_1}_{\rm Q}$--$J^{\Gamma_6}_{\rm Q}$ model, which exhibits similar multiple-$q$ classical ground states to those found in the $J^{\Gamma_1}_{\rm Q}$--$J^{\Gamma_2}_{\rm M}$ model in Sec.~\ref{sec: model1}. 
The model Hamiltonian is given by
\begin{align}
\label{eq: Ham2}
\mathcal{H}= &-\sideset{}{'}{\sum}_{\bm{q}} \{
J^{\Gamma_1}_{\rm Q} Q_u(\bm{q})Q_u(-\bm{q}) \nonumber \\
&+ J^{\Gamma_6}_{\rm Q}\left[Q_{v}(\bm{q})Q_{v}(-\bm{q}) +Q_{xy}(\bm{q})Q_{xy}(-\bm{q}) \right]\},
\end{align}
where $J^{\Gamma_2}_{\rm M}= J^{\Gamma_5}_{\rm M}=J^{\Gamma_5}_{\rm Q}=0$. 
We set $ J^{\Gamma_6}_{\rm Q}=J \cos \theta_{\rm II}$ and $J^{\Gamma_1}_{\rm Q}=J \sin \theta_{\rm II}$ with $J=1$ and $0 \leq \theta_{\rm II} \leq \pi/2$.

\begin{figure}[t!]
\begin{center}
\includegraphics[width=1.0 \hsize ]{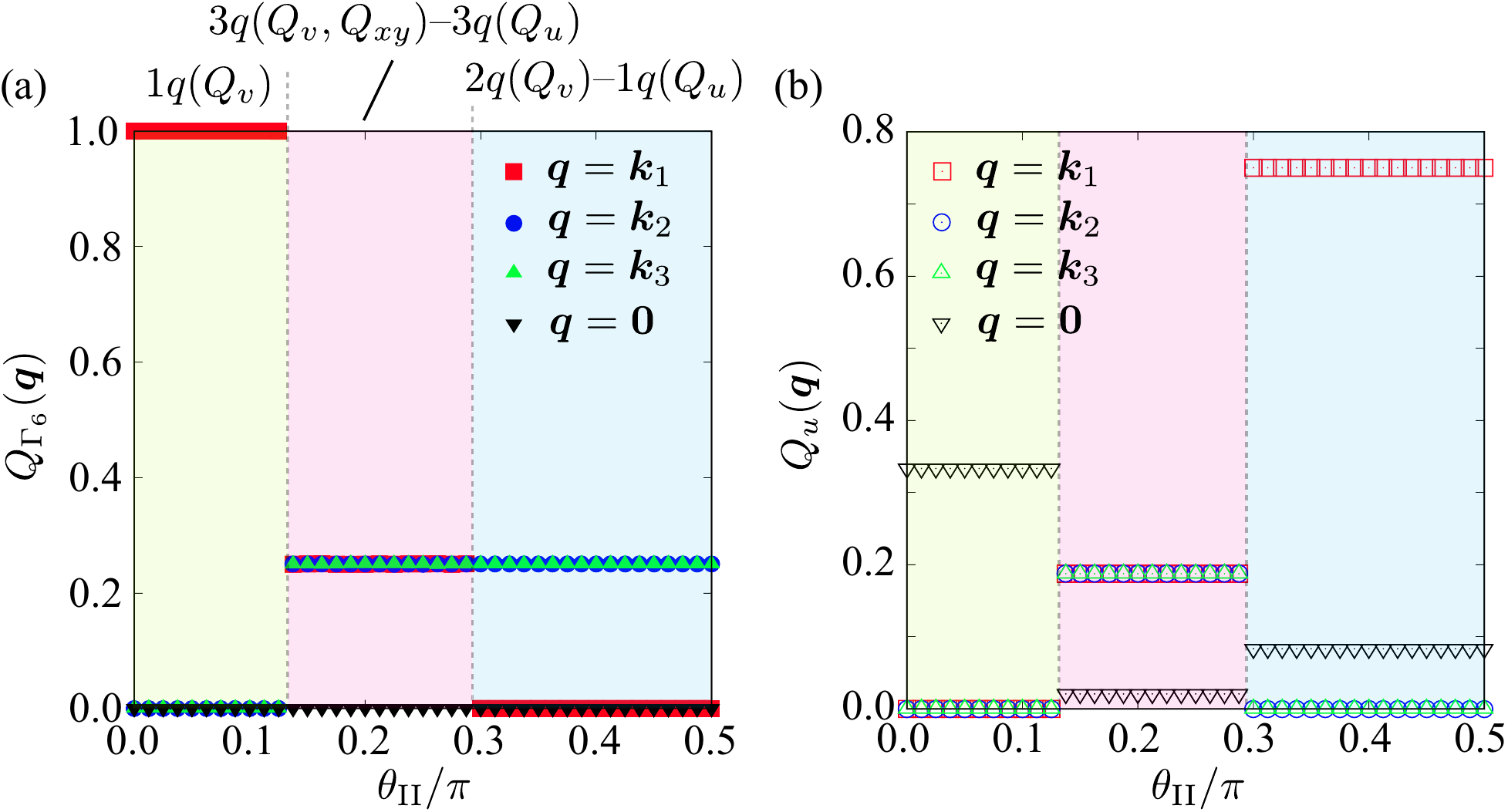} 
\caption{
\label{fig: QuQvQxy}
$\theta_{\rm II}$ dependence of (a) $[Q_{\Gamma_6}(\bm{q})]^2$ and (b) $[Q_u(\bm{q})]^2$ at $\bm{q}=\bm{k}_1$, $\bm{k}_2$, $\bm{k}_3$, and $\bm{0}$ for the model (\ref{eq: Ham2}). 
The vertical dashed lines represent the phase boundaries. 
}
\end{center}
\end{figure}

\begin{figure}[t!]
\begin{center}
\includegraphics[width=1.0 \hsize ]{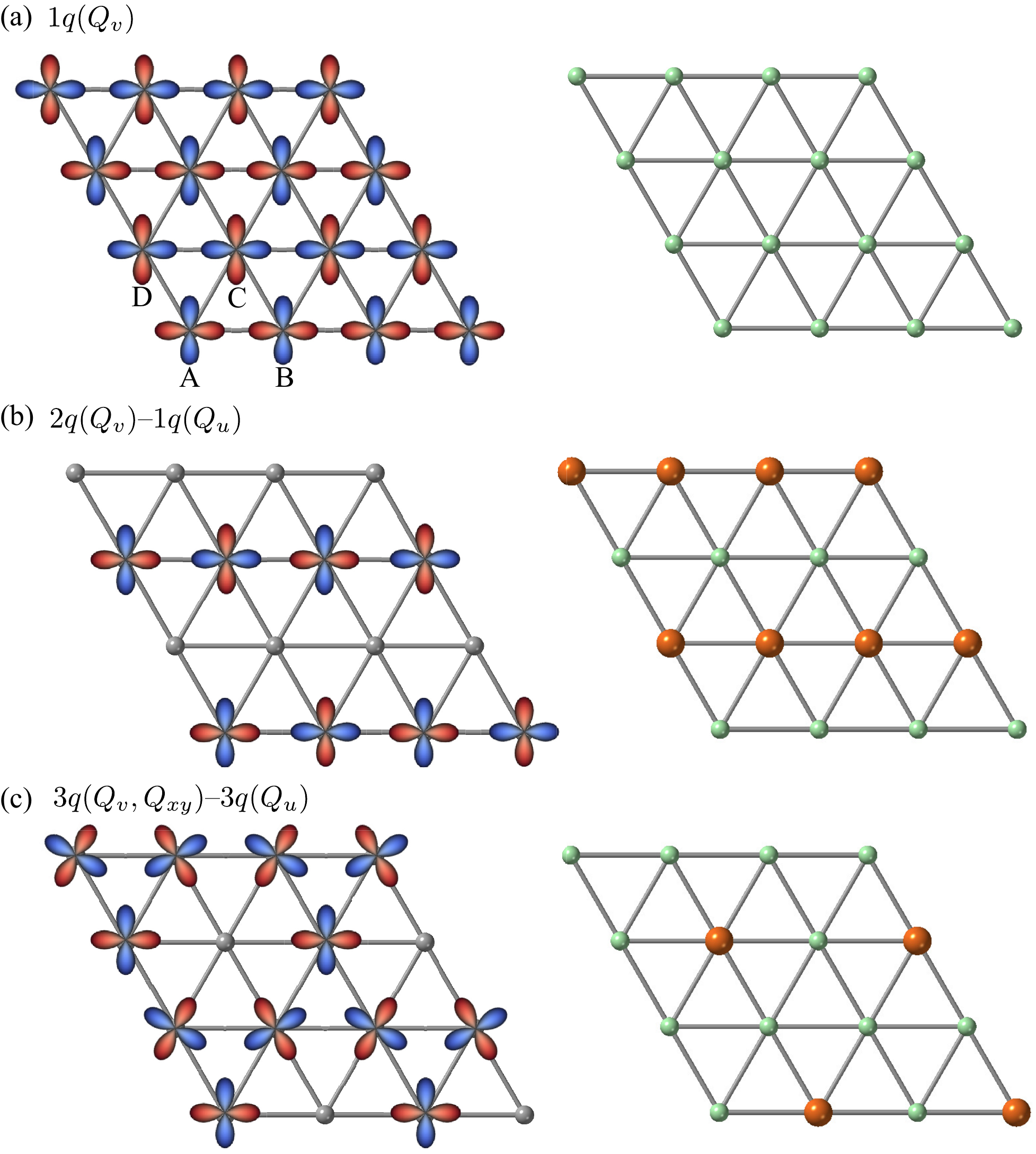} 
\caption{
\label{fig: QuQvQxy_RS}
Schematic real-space configurations of (left panel) $(Q^{\bm{r}}_v, Q^{\bm{r}}_{xy})$ and (right panel) $Q^{\bm{r}}_u$ for (a) the 1$q(Q_v)$, (b) $2q(Q_v)$--$1q(Q_u)$, and (c) $3q(Q_v, Q_{xy})$--$3q(Q_u)$ phases. 
In the left panel, the orbitals represent $(Q^{\bm{r}}_{v}, Q^{\bm{r}}_{xy})$. 
In the right panel, the orange (green) spheres represent $Q^{\bm{r}}_u=2/\sqrt{3}$ ($Q^{\bm{r}}_u=-1/\sqrt{3}$). 
}
\end{center}
\end{figure}

Similarly to the result for the $J^{\Gamma_1}_{\rm Q}$--$J^{\Gamma_2}_{\rm M}$ model, the model (\ref{eq: Ham2}) shows three different phases, as shown in Figs.~\ref{fig: QuQvQxy}(a) and \ref{fig: QuQvQxy}(b); we set $Q_{\Gamma_6}(\bm{q})=\sqrt{[Q_v(\bm{q})]^2+[Q_{xy}(\bm{q})]^2}$. 
For small $\theta_{\rm II}$, the 1$q(Q_v)$ state characterized by $Q_v (\bm{k}_1)=1$ and $[Q_u (\bm{0})]^2=1/3$ is stabilized. 
This state corresponds to the 1$q(M_z)$ state in Sec.~\ref{sec: model1} by replacing $M_z$ with $Q_v$; $\phi=\pi/2$ or $\phi=0$ is selected instead of $\phi=\pi/4$ in Eqs.~(\ref{eq: Mx_coh})--(\ref{eq: Qxy_coh}) at each lattice site. 
The real-space configurations of $Q^{\bm{r}}_v$ and $Q^{\bm{r}}_u$ are shown in the left and right panels of Fig.~\ref{fig: QuQvQxy_RS}(a), respectively. 
It is noted that the state with $Q_{xy} (\bm{k}_1)=1$ and $[Q_u (\bm{0})]^2=1/3$ has the same energy as the 1$q(Q_v)$ state, where such degeneracy is lifted by introducing the two-quadrupole interaction between $Q_v(\bm{q})$ and $Q_{xy}(\bm{q})$ in the forms of $Q_v(\bm{q})Q_v(-\bm{q})-Q_{xy}(\bm{q})Q_{xy}(-\bm{q})$ and $Q_{v}(\bm{q})Q_{xy}(-\bm{q})+Q_{v}(-\bm{q})Q_{xy}(\bm{q})$ or three-quadrupole interaction $Q_v(\bm{0})[3Q_{xy}(\bm{q})Q_{xy}(-\bm{q})-Q_{v}(\bm{q})Q_{v}(-\bm{q})]$~\cite{ishitobi2022triple}.  
For large $\theta_{\rm II}$, the $2q(Q_v)$--$1q(Q_u)$ state is stabilized, which is similar to the $2q(M_z)$--$1q(Q_u)$ state in Sec.~\ref{sec: model1}; the value of $\phi$ is different from each other while $\theta$ is the same. 
The real-space configurations of $Q^{\bm{r}}_v$ and $Q^{\bm{r}}_u$ are shown in the left and right panels of Fig.~\ref{fig: QuQvQxy_RS}(b), respectively. 

Meanwhile, the intermediate state denoted as the 3$q(Q_v, Q_{xy})$--3$q(Q_u)$ state shows a different feature from the $3q'(M_z)$--$3q(Q_u)$ state in Sec.~\ref{sec: model1}. 
The component of $Q_{\Gamma_6}(\bm{q})$ exhibits the isotropic triple-$q$ structure as well as that of $Q_u(\bm{q})$. 
In the real-space configuration of $(Q^{\bm{r}}_v, Q^{\bm{r}}_{xy})$ in the left panel of Fig.~\ref{fig: QuQvQxy_RS}(c), one finds that they form the 120$^{\circ} $ structure of $(Q^{\bm{r}}_v, Q^{\bm{r}}_{xy})$ on the kagome network consisting of the sublattices A, C, and D, i.e., $(Q^{\rm A}_v, Q^{\rm A}_{xy})=(1,0)$, $(Q^{\rm C}_v, Q^{\rm C}_{xy})=(-1/2,-\sqrt{3}/2)$, and $(Q^{\rm D}_v, Q^{\rm D}_{xy})=(-1/2,\sqrt{3}/2)$. 
This means that the two-component $\Gamma_6$ representation is essential to induce the 3$q(Q_v, Q_{xy})$--3$q(Q_u)$ state. 
The configuration of $Q^{\bm{r}}_u$ in the real space is similar to that in the $3q'(M_z)$--$3q(Q_u)$ state, as compared in the right panel of Figs.~\ref{fig: QuMz_RS}(c) and \ref{fig: QuQvQxy_RS}(c)

The energy in each phase is evaluated as 
\begin{align}
E^{1q(Q_v)} &= -J^{\Gamma_6}_{\rm Q}, \\
E^{3q(Q_v, Q_{xy})-3q(Q_u)} &=-\frac{3}{4}J^{\Gamma_6}_{\rm Q}-\frac{9}{16}J^{\Gamma_1}_{\rm Q},  \\
E^{2q(Q_v)-1q(Q_u)} &= -\frac{1}{2}J^{\Gamma_6}_{\rm Q} - \frac{3}{4}J^{\Gamma_1}_{\rm Q}. 
\end{align}
Thus, the phase boundary between the 1$q(Q_v)$ and 3$q(Q_v, Q_{xy})$--3$q(Q_u)$ states is given by $\theta^{(1)}_{\rm II}=\arctan(4/9)\sim 0.133 \pi$ and that between the 3$q(Q_v, Q_{xy})$--3$q(Q_u)$ and $2q(Q_v)$--$1q(Q_u)$ states is given by $\theta^{(2)}_{\rm II}=\arctan(4/3)\sim 0.295 \pi$. 

Let us comment on the anisotropic limit of the quadrupole sector $(Q_v,Q_{xy})$. 
When the interaction for either $Q_v$ or $Q_{xy}$ is neglected, 3$q(Q_v, Q_{xy})$--3$q(Q_u)$ state is replaced by a up-up-down structure of $Q_v$ or $Q_{xy}$. Since this anisotropic limit corresponds to the model discussed in Sec.~\ref{sec: model1}, the energy of this state is given by $-11/16J^{\Gamma_6}_{\rm Q}-9/16J^{\Gamma_1}_{\rm Q}$. See Eq.~(\ref{eq:E3qMz-3qQu}). As a result, $\theta^{(1)}_{\rm II}$ ($\theta^{(2)}_{\rm II}$) moves toward the larger (smaller) $\theta_{\rm II}$. The shift of the phase boundaries is naturally understood from the fact that the energy for the 120$^{\circ}$ structure consisting of the two-dimensional irreducible representation is lower than that for the up-up-down structure consisting of the one-dimensional one.

\subsection{$J^{\Gamma_5}_{\rm Q}$--$J^{\Gamma_6}_{\rm Q}$ model}
\label{sec: model3}

We now discuss the multiple-$q$ instability categorized into 3$q$-II in Table~\ref{tab: multiQ}. 
We first consider the $J^{\Gamma_5}_{\rm Q}$--$J^{\Gamma_6}_{\rm Q}$ model, which is given by 
\begin{align}
\label{eq: Ham3}
\mathcal{H}= &-\sideset{}{'}{\sum}_{\bm{q}} \{
J^{\Gamma_5}_{\rm Q} \left[Q_{yz}(\bm{q})Q_{yz}(-\bm{q}) +Q_{zx}(\bm{q})Q_{zx}(-\bm{q}) \right]
 \nonumber \\
 &+ J^{\Gamma_6}_{\rm Q}\left[Q_{v}(\bm{q})Q_{v}(-\bm{q}) +Q_{xy}(\bm{q})Q_{xy}(-\bm{q}) \right]
 \},
\end{align}
where $J^{\Gamma_2}_{\rm M}= J^{\Gamma_5}_{\rm M}=J^{\Gamma_1}_{\rm Q}=0$. 
We set $ J^{\Gamma_5}_{\rm Q}=J \cos \theta_{\rm III}$ and $J^{\Gamma_6}_{\rm Q}=J \sin \theta_{\rm III}$ with $J=1$ and $0 \leq \theta_{\rm III} \leq \pi/2$.

\begin{figure}[t!]
\begin{center}
\includegraphics[width=1.0 \hsize ]{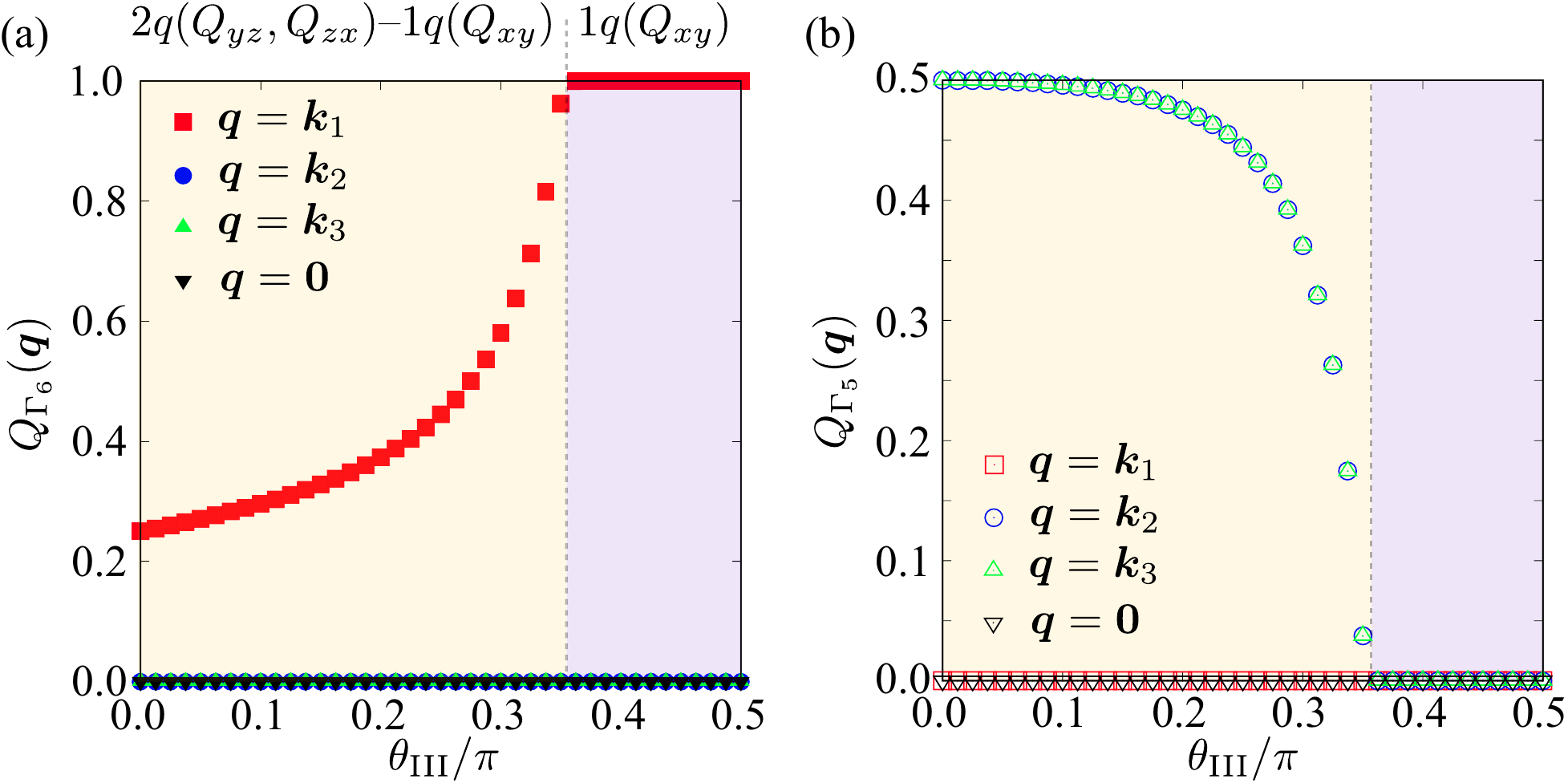} 
\caption{
\label{fig: QvQxyQyzQzx}
$\theta_{\rm III}$ dependence of (a) $[Q_{\Gamma_5}(\bm{q})]^2$ and (b) $[Q_{\Gamma_6}(\bm{q})]^2$ at $\bm{q}=\bm{k}_1$, $\bm{k}_2$, $\bm{k}_3$, and $\bm{0}$ for the model (\ref{eq: Ham3}). 
The vertical dashed lines represent the phase boundaries. 
}
\end{center}
\end{figure}

\begin{figure}[t!]
\begin{center}
\includegraphics[width=1.0 \hsize ]{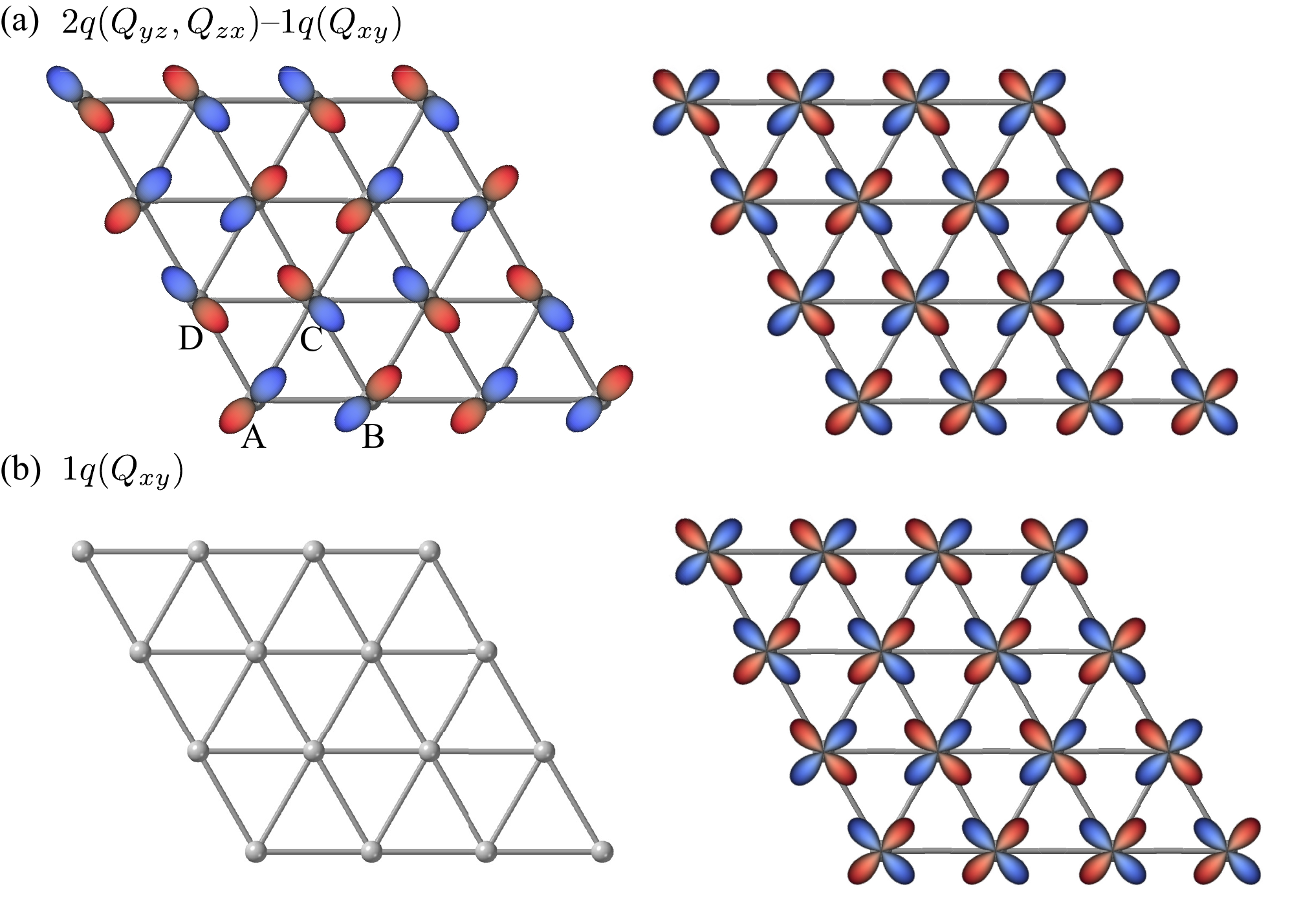} 
\caption{
\label{fig: QvQxyQyzQzx_RS}
Schematic real-space configurations of (left panel) $(Q^{\bm{r}}_{yz}, Q^{\bm{r}}_{zx})$ and (right panel) $(Q^{\bm{r}}_v, Q^{\bm{r}}_{xy})$ for (a) the $2q(Q_{yz}, Q_{zx})$--$1q(Q_{xy})$ and (b) $1q(Q_{xy})$ phases. 
In the left panel, the orbitals represent $(Q^{\bm{r}}_{yz}, Q^{\bm{r}}_{zx})$. 
In the right panel, the orbitals represent $(Q^{\bm{r}}_{v}, Q^{\bm{r}}_{xy})$.
}
\end{center}
\end{figure}

Figures~\ref{fig: QvQxyQyzQzx}(a) and \ref{fig: QvQxyQyzQzx}(b) show the behavior of $[Q_{\Gamma_5}(\bm{q})]^2$ and $[Q_{\Gamma_6}(\bm{q})]^2$ against $\theta_{\rm III}$, respectively; we set $Q_{\Gamma_5}(\bm{q})=\sqrt{[Q_{yz}(\bm{q})]^2+[Q_{zx}(\bm{q})]^2}$.  
There are two phases as $\theta_{\rm III}$ varies: the $2q(Q_{yz}, Q_{zx})$--$1q(Q_{xy})$ state for small $\theta_{\rm III}$ and the $1q(Q_{xy})$ state for large $\theta_{\rm III}$. 
The former state is characterized by the triple-$q$ state, while the latter is by the single-$q$ state. 
In contrast to the cases categorized into the 3$q$-I in Secs.~\ref{sec: model1} and \ref{sec: model2}, the phase transition between the $2q(Q_{yz}, Q_{zx})$--$1q(Q_{xy})$ and $1q(Q_{xy})$ states is of second order. 

In the $2q(Q_{yz}, Q_{zx})$--$1q(Q_{xy})$ state for small $\theta_{\rm III}$, the double-$q$ structure appears in $Q_{\Gamma_5}(\bm{q})$, while the single-$q$ structure appears in $Q_{\Gamma_6}(\bm{q})$, as shown in Figs.~\ref{fig: QvQxyQyzQzx}(a) and \ref{fig: QvQxyQyzQzx}(b). 
Specifically, this state is characterized by nonzero $Q_{xy}(\bm{k}_1)$ and $Q_{yz}(\bm{k}_2)=Q_{zx}(\bm{k}_3)$, whose real-space configurations of $(Q^{\bm{r}}_{yz}, Q^{\bm{r}}_{zx})$ and $Q^{\bm{r}}_{xy}$ are shown in the left and right panels of Fig.~\ref{fig: QvQxyQyzQzx_RS}(a), respectively.
It is noted that the state with nonzero $Q_{v}(\bm{k}_1)$ instead of $Q_{xy}(\bm{k}_1)$ gives the same energy, where $(Q^{\bm{r}}_{yz}, Q^{\bm{r}}_{zx})$ are rotated by $\pi/4$ around the $z$ axis. 
As only the single-$q$ state is realized in the $J^{\Gamma_5}_{\rm Q}$ model when $\theta_{\rm III}=0$ as shown in Table~\ref{tab: multiQ}, the appearance of the triple-$q$ state is attributed to the presence of $J^{\Gamma_6}_{\rm Q}$. 
Indeed, nonzero $Q_{xy}(\bm{k}_1)$ appears for infinitesimally small $\theta_{\rm III}$. 

The interplay between $[Q_{yz}(\bm{q}),Q_{zx}(\bm{q})]$ and $Q_{xy}(\bm{q})$ is understood from the real-space multipole expressions (\ref{eq: Mx_coh})--(\ref{eq: Qxy_coh}). 
To maximize the three components of $Q^{\bm{r}}_{yz}$, $Q^{\bm{r}}_{zx}$, and $Q^{\bm{r}}_{xy}$, it is desirable to satisfy $M^{\bm{r}}_x=M^{\bm{r}}_y=M^{\bm{r}}_z=Q^{\bm{r}}_{v}=0$ due to the local constraint of the multipole length. 
Then, one finds $\phi=\pi/4$ and $\alpha_1, \alpha_2=0$ or $\pi$, where the expressions (\ref{eq: Qu_coh}) and (\ref{eq: Qyz_coh})--(\ref{eq: Qxy_coh}) are rewritten as 
\begin{align}
\label{eq: Qu_coh2}
Q^{\bm{r}}_u&=\frac{1}{2\sqrt{3}}(1+3\cos 2\theta), \\
\label{eq: Qyz_coh2}
Q^{\bm{r}}_{yz}&= \pm \frac{1}{\sqrt{2}} \sin 2\theta, \\
\label{eq: Qzx_coh2}
Q^{\bm{r}}_{zx}&= \pm \frac{1}{\sqrt{2}} \sin 2\theta, \\
\label{eq: Qxy_coh2}
Q^{\bm{r}}_{xy}&= \pm  \sin^2 \theta. 
\end{align}
To be specific, $(Q^{\rm A}_{yz}, Q^{\rm A}_{zx}, Q^{\rm A}_{xy})=(-\sin 2\theta/\sqrt{2}, -\sin 2\theta/\sqrt{2}, \sin^2\theta)$, $(Q^{\rm B}_{yz}, Q^{\rm B}_{zx}, Q^{\rm B}_{xy})=(\sin 2\theta/\sqrt{2}, \sin 2\theta/\sqrt{2}, \sin^2\theta)$, $(Q^{\rm C}_{yz}, Q^{\rm C}_{zx}, Q^{\rm C}_{xy})=(\sin 2\theta/\sqrt{2}, -\sin 2\theta/\sqrt{2}, -\sin^2\theta)$, and $(Q^{\rm D}_{yz}, Q^{\rm D}_{zx}, Q^{\rm D}_{xy})=(-\sin 2\theta/\sqrt{2}, \sin 2\theta/\sqrt{2}, -\sin^2\theta)$ in Fig.~\ref{fig: QvQxyQyzQzx_RS}(a). 
Thus, $Q^{\bm{r}}_{xy} \neq 0$ when $Q^{\bm{r}}_{yz}$ and $Q^{\bm{r}}_{zx} \neq 0$ except for $\theta=\pi/2$, which contributes to the energy by $J^{\Gamma_6}_{\rm Q}$. 
Moreover, the expressions (\ref{eq: Qu_coh2})--(\ref{eq: Qxy_coh2}) indicate that $Q_{yz}^{\bm{r}}$ and $Q_{zx}^{\bm{r}}$ are the same magnitude at each site; the orbital at each lattice site seen from the positive $z$-axis extends to the direction tilted by $45^{\circ}$ or $135^{\circ}$ from the $x$ axis, as shown in the left panel of Fig.~\ref{fig: QvQxyQyzQzx_RS}(a).
When considering the modulation of $Q_{xy}(\bm{k}_1)$, the double-$q$ superposition of $Q_{yz}(\bm{k}_2)=Q_{zx}(\bm{k}_3)$ gives the lowest energy.
This is also naively understood from the fact that there is an effective coupling in the form of $Q_{xy}(\bm{k}_1)Q_{yz}(\bm{k}_2)Q_{zx}(\bm{k}_3)$.
Similarly, one finds the coupling between $[Q_{yz}(\bm{q}),Q_{zx}(\bm{q})]$ and $Q_{v}(\bm{q})$ when setting $\phi=\pi/2$ or $0$ instead of $\phi=\pi/4$. 

When $\theta_{\rm III}$ increases, the magnitudes of $Q_{yz}(\bm{q})$ and $Q_{zx}(\bm{q})$ become smaller and they vanish at $\theta_{\rm III} \simeq 0.356 \pi $, which means the transition to the $1q(Q_{xy})$ state. 
This state has the same energy as the 1$q(Q_v)$ state does in Sec.~\ref{sec: model2}, whose configurations of $(Q^{\bm{r}}_{yz}, Q^{\bm{r}}_{zx})$ and $Q^{\bm{r}}_{xy}$ are shown in the left and right panels of Fig.~\ref{fig: QvQxyQyzQzx_RS}(b), respectively.

\subsection{$J^{\Gamma_5}_{\rm M}$--$J^{\Gamma_6}_{\rm Q}$ model}
\label{sec: model4}

Finally, let us consider the $J^{\Gamma_5}_{\rm M}$--$J^{\Gamma_6}_{\rm Q}$ model, which is given by 
\begin{align}
\label{eq: Ham4}
\mathcal{H}= &-\sideset{}{'}{\sum}_{\bm{q}} \{
J^{\Gamma_5}_{\rm M} \left[M_{x}(\bm{q})M_{x}(-\bm{q}) +M_{y}(\bm{q})M_{y}(-\bm{q}) \right]
 \nonumber \\
 &+ J^{\Gamma_6}_{\rm Q}\left[Q_{v}(\bm{q})Q_{v}(-\bm{q}) +Q_{xy}(\bm{q})Q_{xy}(-\bm{q}) \right]
 \},
\end{align}
where $J^{\Gamma_2}_{\rm M}= J^{\Gamma_1}_{\rm Q}=J^{\Gamma_5}_{\rm Q}=0$. 
We set $ J^{\Gamma_5}_{\rm M}=J \cos \theta_{\rm IV}$ and $J^{\Gamma_6}_{\rm Q}=J \sin \theta_{\rm IV}$ with $J=1$ and and $0 \leq \theta_{\rm IV} \leq \pi/2$.

\begin{figure}[t!]
\begin{center}
\includegraphics[width=1.0 \hsize ]{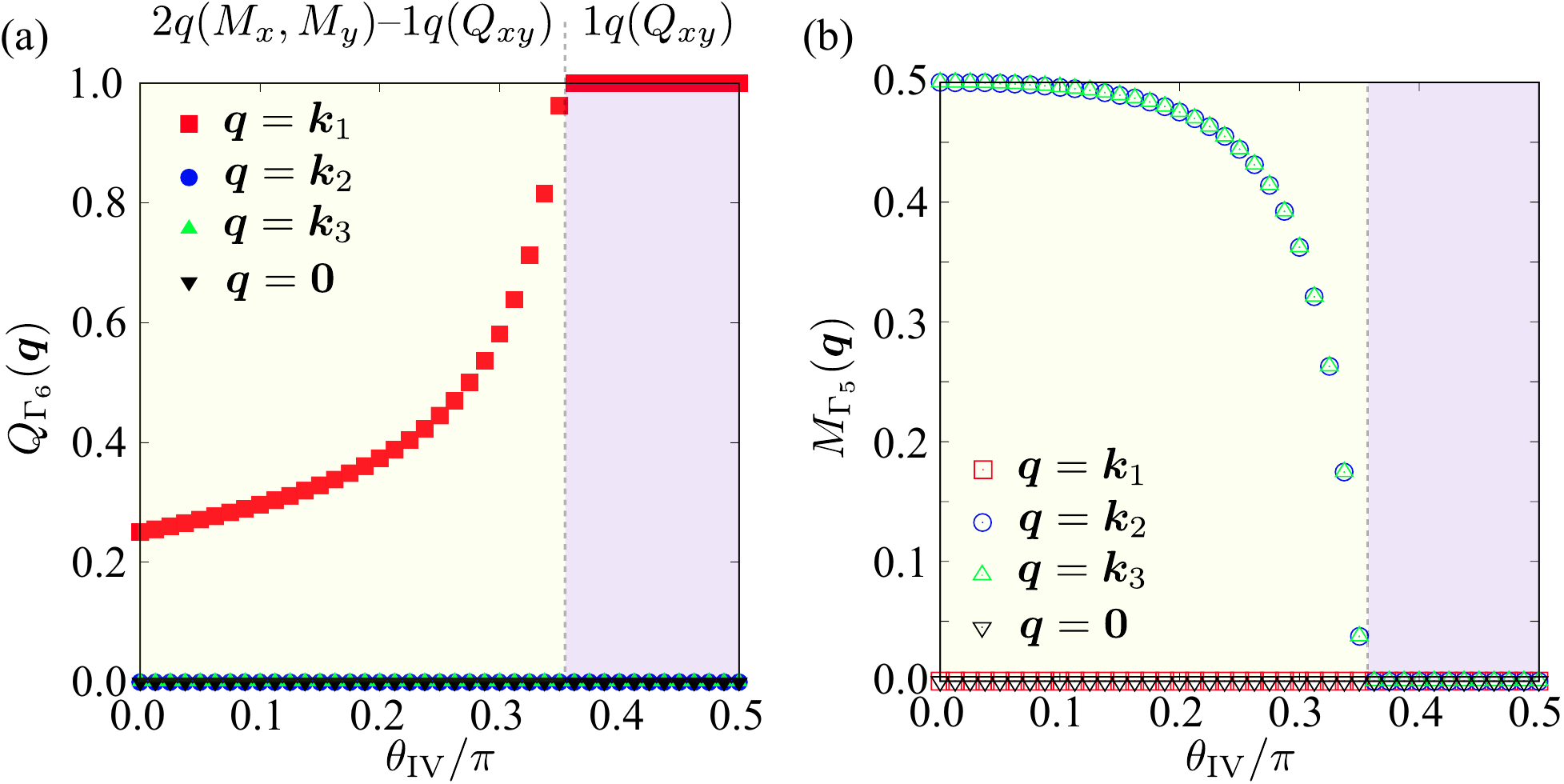} 
\caption{
\label{fig: QvQxyMxMy}
$\theta_{\rm IV}$ dependence of (a) $[Q_{\Gamma_6}(\bm{q})]^2$ and (b) $[M_{\Gamma_5}(\bm{q})]^2$ at $\bm{q}=\bm{k}_1$, $\bm{k}_2$, $\bm{k}_3$, and $\bm{0}$ for the model (\ref{eq: Ham4}). 
The vertical dashed lines represent the phase boundaries. 
}
\end{center}
\end{figure}

\begin{figure}[t!]
\begin{center}
\includegraphics[width=1.0 \hsize ]{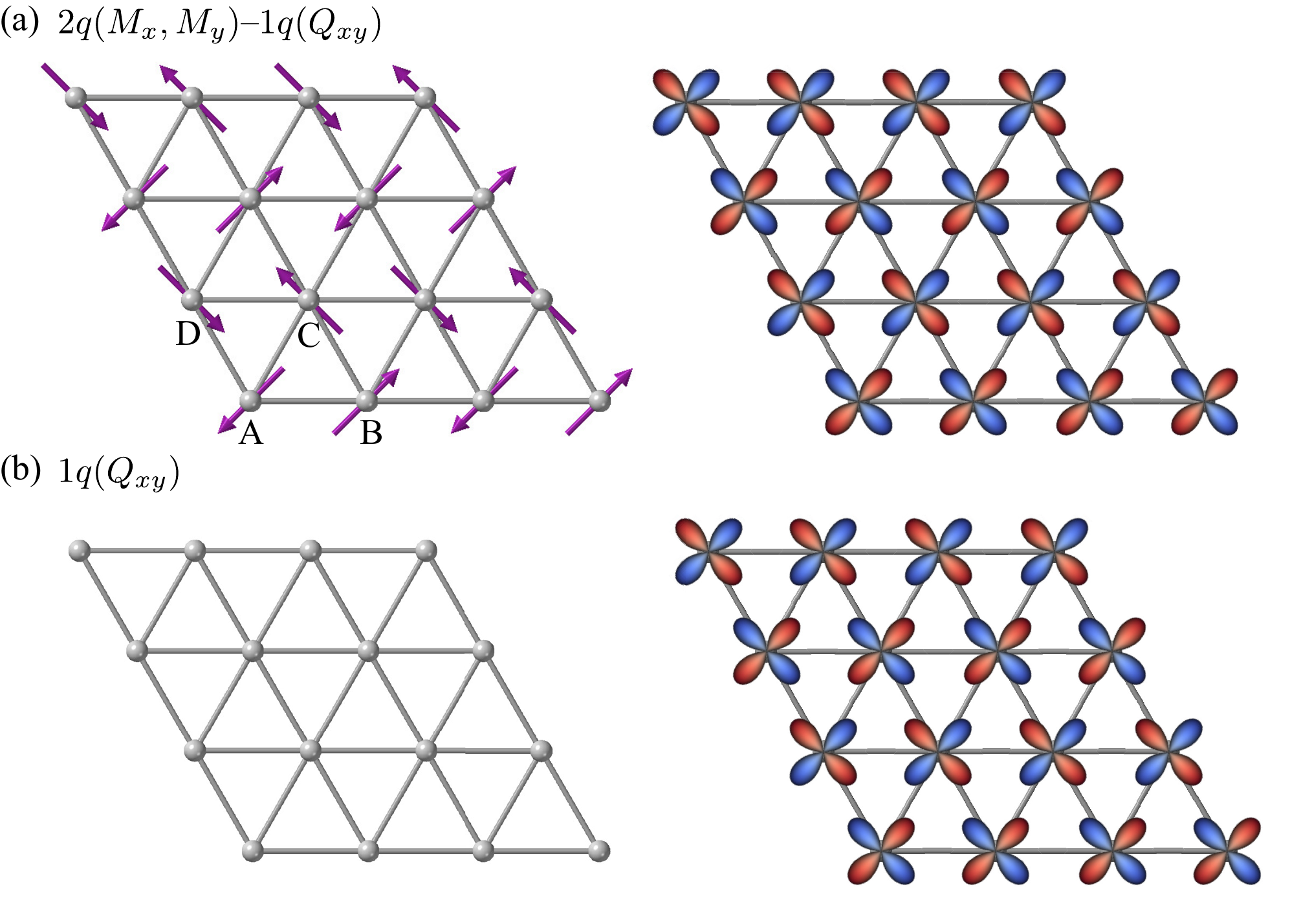} 
\caption{
\label{fig: QvQxyMxMy_RS}
Schematic real-space configurations of (left panel) $(M^{\bm{r}}_{x}, M^{\bm{r}}_{y})$ and (right panel) $(Q^{\bm{r}}_v, Q^{\bm{r}}_{xy})$ for (a) the $2q(M_{x}, M_{y})$--$1q(Q_{xy})$ and (b) $1q(Q_{xy})$ phases. 
In the left panel, the spins represent $(M^{\bm{r}}_{x}, M^{\bm{r}}_{y})$. 
In the right panel, the orbitals represent $(Q^{\bm{r}}_{v}, Q^{\bm{r}}_{xy})$.
}
\end{center}
\end{figure}

Figures~\ref{fig: QvQxyMxMy}(a) and \ref{fig: QvQxyMxMy}(b) show the $\theta_{\rm IV}$ dependence of $[Q_{\Gamma_6}(\bm{q})]^2$ and $[M_{\Gamma_5}(\bm{q})]^2$, respectively; we set $M_{\Gamma_5}(\bm{q})=\sqrt{[M_{x}(\bm{q})]^2+[M_{y}(\bm{q})]^2}$.  
The results in Figs.~\ref{fig: QvQxyMxMy}(a) and \ref{fig: QvQxyMxMy}(b) are similar to those in Figs.~\ref{fig: QvQxyQyzQzx}(b) and \ref{fig: QvQxyQyzQzx}(a) by replacing $[M_x(\bm{q}), M_y (\bm{q})]$ with $[Q_{zx}(\bm{q}), Q_{yz} (\bm{q})]$. 
In the real-space picture, this mapping is obtained by using $\alpha_1, \alpha_2=\pi/2$, $3\pi/2$ instead of $\alpha_1, \alpha_2=0$, $\pi$ in Eqs.~(\ref{eq: Mx_coh}), (\ref{eq: My_coh}), (\ref{eq: Qyz_coh}), and (\ref{eq: Qzx_coh}) while keeping $\phi=\pi/4$ for nonzero $Q^{\bm{r}}_{xy}$ or $\phi=\pi/2$, 0 for nonzero $Q^{\bm{r}}_{v}$; we show the real-space pictures of $(M^{\bm{r}}_{x}, M^{\bm{r}}_{y})$ and $(Q^{\bm{r}}_v, Q^{\bm{r}}_{xy})$ for $\phi=\pi/4$ in the left and right panels of Fig.~\ref{fig: QvQxyMxMy_RS}(a), respectively. 
This state is the ground state for $\theta_{\rm IV}\lesssim 0.356 \pi$. 
For $\theta_{\rm IV} \gtrsim 0.356 \pi$, the 1$q(Q_{xy})$ or 1$q(Q_v)$ state is realized. 
Their configurations of $(M^{\bm{r}}_{x}, M^{\bm{r}}_{y})$ and $(Q^{\bm{r}}_v, Q^{\bm{r}}_{xy})$ are shown in Fig.~\ref{fig: QvQxyMxMy_RS}(b).

\section{Discussion}
\label{sec: Discussion}

In this section, we first comment on the qualitative understanding of multiple-$q$ multipole orders presented in Sec.~\ref{sec: model1}, and then discuss the effect of the hexagonal crystalline electric field on the phases discussed in Sec.~\ref{sec: Delta} and the appearance of a charge modulation when additionally taking into account the charge degree of freedom in Sec.~\ref{sec: Charge disproportionation}.

\subsection{Qualitative understanding of the multiple-$q$ multipole orders}
\label{sec: mechanism}
We have discussed four types of minimal models in Sec.~\ref{sec: Results}, which possess various triple-$q$ ground states in their phase diagrams. 
They can be categorized into two groups. 
One is the model which includes the $Q_u$ degree of freedom as analyzed in Secs.~\ref{sec: model1} and \ref{sec: model2}. 
The other is those without $Q_u$ discussed in Secs.~\ref{sec: model3} and \ref{sec: model4}. 
As listed in Table~\ref{tab: multiQ}, the former is represented by 3$q$-I, while the latter by 3$q$-II. The difference between the two groups lies in the presence of the 3$q$-3$q$-type triple-$q$ orders in their ground states, while the 1$q$ and 2$q$-1$q$ types are present in both groups. See Figs.~\ref{fig: QuMz},~\ref{fig: QuQvQxy},~\ref{fig: QvQxyQyzQzx}, and \ref{fig: QvQxyMxMy}.

Let us discuss these aspects from the point of view of the Landau free energy at finite temperatures. Except for $Q_u^{\bm r}$, all the multipole matrices $X^{\bm r}$ in Eqs.~(\ref{eq: Mxmat})-(\ref{eq: Qxymat}) satisfy the relation $(X^{\bm r})^3=X^{\bm r}$. This readily means that there is no cubic potential $\propto (X^{\bm r})^3$ in their local free energy even when $X^{\bm r}$ is even under the time-reversal operation. Thus, for realizing triple-$q$ orders within a single multipole sector one needs a particular fourth-order term originating from e.g., higher-order Fermi surface nestings~\cite{Akagi_PhysRevLett.108.096401, Hayami_PhysRevB.90.060402, Ozawa_doi:10.7566/JPSJ.85.103703, Hayami_PhysRevB.95.224424}. However such processes are not considered in this paper. In contrast, when $X^{\bm r}=Q_u^{\bm r}$, one finds $(Q^{\bm r}_u)^3\ne Q^{\bm r}_u$. Thus, the local Landau free energy possesses the cubic term $\propto (Q^{\bm r}_u)^3$. This trivially contains the coupling $Q_u(\bm{k}_1)Q_u(\bm{k}_2)Q_u(\bm{k}_3)$, which favors the 3$q$($Q_u$) state \cite{tsunetsugu2021quadrupole}. Once the triple-$q$ order takes place in the $Q_u$ sector, it naturally couples with $M_z({\bm k}_{1,2,3})$ and $Q_{v,xy}({\bm k}_{1,2,3})$ as discussed in Sec.~\ref{sec: model1} and Sec.~\ref{sec: model2}, respectively.

At zero temperature, the 3$q$($Q_u$) state is unstable since this cannot satisfy the local constraint $\sum_X (X^{\bm{r}})^2 = 4/3$. 
In other words, the smaller length of $|Q_u^{\bm r}|<\sqrt{3}/2$ should be compensated by the other multipoles. 
This situation is also the case for the other phases. 
For example, 1$q(Q_{xy})$ with $\theta=\pi/2$ or $3\pi/2$ in Eq.~(\ref{eq: Qu_coh}) accompanies with the uniform $Q_u$, which is not as shown in Fig.~\ref{fig: QvQxyMxMy_RS}(b). As lowering $\theta_{\rm IV}$ i.e., increasing $J_{\rm M}^{\Gamma_5}$, there is a critical $\theta_{\rm IV}$ at which the $M_{x,y}$ order takes place. 
Concerning the stability of 3$q$-3$q$ type phases against the increase in $J_{\rm Q}^{\Gamma_1}$ i.e., $\theta_{\rm I,II}$, the numerical results show that 2$q$-1$q$ type orders overwhelm for $\theta_{\rm I, II} \sim \pi/2$ both in the two models in Secs.~\ref{sec: model1} and \ref{sec: model2}. 
This is quite natural since we have already known the stable 1$q$ state for a single multipole model as summarized in Table~ \ref{tab: multiQ}. 
Thus, the 3$q$-3$q$ type phases have a change to appear only when the two interactions $J_{\rm Q}^{\Gamma_1}$ and $J_{\rm M}^{\Gamma_2}(J_{\rm Q}^{\Gamma_6})$ compete.

\subsection{Hexagonal crystalline electric field}
\label{sec: Delta}

We have so far investigated the multiple-$q$ instability at $\Delta=0$ for the four models in Secs.~\ref{sec: model1}--\ref{sec: model4}. 
Here, we briefly discuss the effect of $\Delta$, which corresponds to the hexagonal crystalline electric field, on the phase diagrams. 
It turns out that, even for finite $\Delta$, the results are qualitatively similar to those shown in Figs.~\ref{fig: QuMz}, \ref{fig: QuQvQxy}, \ref{fig: QvQxyQyzQzx}, and \ref{fig: QvQxyMxMy} for all the models, at least up to $|\Delta|/J<0.5$; no new phases appear for $\Delta \neq 0$. 
In the $J^{\Gamma_1}_{\rm Q}$--$J^{\Gamma_2}_{\rm M}$ ($J^{\Gamma_1}_{\rm Q}$--$J^{\Gamma_6}_{\rm Q}$) model, phase boundaries tend to move to the larger $\theta_{\rm I}$ ($\theta_{\rm II}$) for $\Delta>0$; the regions for the 1$q(M_z)$ and $3q'(M_z)$--$3q(Q_u)$ [1$q(Q_v)$ and  $3q(Q_v, Q_{xy})$--$3q(Q_u)$] states become larger. 
Meanwhile, phase boundaries tend to move to the smaller $\theta_{\rm I}$ and $\theta_{\rm II}$ for $\Delta<0$; the regions for the $2q(M_z)$--$1q(Q_u)$ and $2q(Q_v)$--$1q(Q_u)$ states become larger, respectively. 
This tendency is naturally understood from the fact that positive (negative) $\Delta$ favors the state with the small (large) $Q_u(\bm{0})$ component. 
Similarly, in the $J^{\Gamma_5}_{\rm Q}$--$J^{\Gamma_6}_{\rm Q}$ ($J^{\Gamma_5}_{\rm M}$--$J^{\Gamma_6}_{\rm Q}$) model, the $1q(Q_{xy})$ state extends toward the smaller $\theta_{\rm III, IV}$ regime for $\Delta>0$, while shrinks for $\Delta<0$.
This is again owing to $Q^{\bm r}_u$; $Q^{\bm r}_u=-1/\sqrt{3}$ in all the sublattices for the $1q(Q_{xy})$ state.

\subsection{Charge disproportionation}
\label{sec: Charge disproportionation}

Some of the phases discussed in Sec.~\ref{sec: Results} accompany charge disproportionation when one extends the classical spin model (\ref{eq: Ham}) to a model with itinerant electrons, as in the Hubbard model and the Kondo lattice model. 
As the $Q^{\bm{r}}_u$ directly couples to the charge degrees of freedom, the nonuniform distribution of $Q^{\bm{r}}_u$ leads to the charge disproportionation once the itinerant electrons are taken into account. 
In the $J^{\Gamma_1}_{\rm Q}$--$J^{\Gamma_2}_{\rm M}$ model in Sec.~\ref{sec: model1}, the $2q(M_z)$--$1q(Q_u)$ state exhibits the charge disproportionation with its modulation vector $\bm{k}_1$ since the site symmetry for the sublattices A and B is different from that for C and D. 
For the $3q'(M_z)$--$3q(Q_u)$ state, at least 1:3-type charge disproportionation occurs in accord with the $Q_u^{\bm{r}}$ alignment. 
Similarly, the $2q(Q_v)$--$1q(Q_u)$ and $3q(Q_v, Q_{xy})$--$3q(Q_u)$ states in the $J^{\Gamma_1}_{\rm Q}$--$J^{\Gamma_6}_{\rm Q}$ model in Sec.~\ref{sec: model2} exhibit a similar charge disproportionation to that in the $2q(M_z)$--$1q(Q_u)$ and $3q'(M_z)$--$3q(Q_u)$ states, respectively. 
In contrast, no charge disproportionation occurs in the phases in the $J^{\Gamma_5}_{\rm Q}$--$J^{\Gamma_6}_{\rm Q}$ and $J^{\Gamma_5}_{\rm M}$--$J^{\Gamma_6}_{\rm Q}$ models.

\section{Summary}
\label{sec: Summary}

To summarize, we have investigated the multiple-$q$ instability in the $S=1$ model with magnetic dipole and electric quadrupole degrees of freedom on a triangular lattice. 
By systematically performing the simulated annealing for the effective models with different multipole-multipole interactions, we have found that four out of the fifteen models show the triple-$q$ ground states. 
These triple-$q$ states are stabilized as a consequence of the interplay between the dipole and quadrupole interactions, which does not require single-ion magnetic anisotropy and higher-order interactions. 
We have also discussed the effect of the hexagonal crystalline electric field and the appearance of charge disproportionation.
Our results underscore the cooperative effect brought about by the coexistence of the dipole and quadrupole degrees of freedom, which gives rise to rich multiple-$q$ states. 
Since the recent studies suggested the possibility of similar coexisting states of multipole moments in materials, such as PrV$_2$Al$_{20}$~\cite{Ishitobi_PhysRevB.104.L241110} and UNi$_4$B~\cite{ishitobi2022triple}, and the CP$^2$ skyrmion phase in triangular magnets, such as $ABX_3$, $BX_2$, and $AB$O$_2$ ($A$: alkali metal, $B$: transition metal, and $X$: halogen atom)~\cite{collins1997review, mcguire2017crystal, botana2019electronic, zhang2022cp}, our study will stimulate further analyses about exotic states with the quadrupole degree of freedom.

Finally, our effective model can serve as a starting model to investigate minimal conditions to understand the multiple-$q$ multipole instability in other various situations. 
For example, the model can be used for the exploration of the multiple-$q$ instability with long-period modulations, such as the spiral state and skyrmion crystal consisting of the dipole and quadrupole moments, when the interactions between finite ordering vectors are taken into account. 
Furthermore, the model can be straightforwardly extended to those including other higher-rank multipole moments, such as octupole moments, which might lead to further unknown states with multiple multipole moments. 
Analyses about such systems will clarify possible multiple-$q$ states and their associated phase transitions triggered by the multipole-multipole correlations.

\begin{acknowledgments}
This research was supported by JSPS KAKENHI Grants Numbers JP21H01037, JP21H01031, JP22H04468, JP22H00101, JP22H01183, and by JST PRESTO (JPMJPR20L8). 
Parts of the numerical calculations were performed in the supercomputing systems in ISSP, the University of Tokyo.
\end{acknowledgments}

\bibliographystyle{apsrev}
\bibliography{ref}

\end{document}